\documentclass{elsart}

\usepackage{graphicx}
\usepackage{float}
\usepackage{amsmath}

\begin{document}
%%% to have more flexibility in the placement of figures
\renewcommand\floatpagefraction{.85} \renewcommand\topfraction{.85}
\renewcommand\bottomfraction{.85} \renewcommand\textfraction{.15}

\renewcommand\floatpagefraction{.95} \renewcommand\topfraction{.95}
\renewcommand\bottomfraction{.95} \renewcommand\textfraction{.05}

%%%the name of an orbit (permutation is irrelevant)
\newcommand\orbit[1]{\langle #1 \rangle}

%%%the name of a periodic point
\newcommand\perpoint[1]{\overline{#1}}

%%% a symbol
\newcommand\sym[1]{\text{``}#1\text{''}}

%%%topological names
\newcommand\name{\mathcal{N}}
\newcommand\nameof[1]{\name(#1)}
\newcommand\nameofi[2]{\name_{#2}(#1)}

%%% names indicated by partitions
\newcommand\pname{N}
\newcommand\pnameof[2]{N_{#2}(#1)}
\newcommand\pa{\Gamma}

%%% regions of a partition
\newcommand\region{\Delta}

%%% the border of a partition
\newcommand\border{\beta}
\newcommand\borderof[1]{\border(#1)}
\newcommand\borderofi[2]{\border_{#2}(#1)}

%%% total number of orbits
\newcommand\orbnum{m}

\begin{frontmatter}

\title{From template analysis\\ to generating partitions I:\\ Periodic
  orbits, knots and symbolic encodings}

\author{Jérôme Plumecoq \and Marc Lefranc\thanksref{CORR}}

\address{Laboratoire de Physique des Lasers, Atomes, Mol\'ecules, UMR
  CNRS 8523,\\
  Centre d'\'Etudes et de Recherches Lasers et Applications,\\
  Universit\'e de Lille I, F-59655 Villeneuve d'Ascq Cedex, France}
\thanks[CORR] {Corresponding author. E-mail: Marc.Lefranc@univ-lille1.fr}

\begin{abstract}
  We present a detailed algorithm to construct symbolic encodings for
  chaotic attractors of three-dimensional flows. It is based on a
  topological analysis of unstable periodic orbits embedded in the
  attractor and follows the approach proposed by Lefranc {\em et
    al.\/} [\emph{Phys.  Rev.  Lett.} {\bf 73}, 1364 (1994)]. For each
  orbit, the symbolic names that are consistent with its
  knot-theoretic invariants and with the topological structure of the
  attractor are first obtained using template analysis. This
  information, and the locations of the periodic orbits in the section
  plane, are then used to construct a generating partition by means of
  triangulations.  We provide numerical evidence of the validity of
  this method by applying it successfully to sets of more than 1500
  periodic orbits extracted from numerical simulations, and obtain
  partitions whose border is localized with a precision of 0.01\%. A
  distinctive advantage of this approach is that the solution is
  progressively refined using higher-period orbits, which makes it
  robust to noise, and suitable for analyzing experimental time
  series. Furthermore, the resulting encodings are by construction
  consistent in the corresponding limits with those rigorously known
  for both one-dimensional and hyperbolic maps.
\end{abstract}
\end{frontmatter}

\tableofcontents
\newpage
\section{Introduction}
\label{sec:intro}

Symbolic dynamics is a powerful approach to chaotic dynamics. It
consists in representing trajectories in a chaotic attractor by
sequences of symbols from a finite alphabet, in a way that preserves
the essential properties of the
dynamics~\cite{Guckenheimer83book,Katok95a,Hao89a,Badii97a}. It is not
only central to some of the most fundamental theorems of dynamical
systems theory (see, e.g., \cite{Guckenheimer83book,Katok95a}), but
can also be of utmost importance with a view to practical
applications, such as for transmitting numeric streams over chaotic
signals~\cite{Hayes93a,Hayes94a}.

However, we currently have a good understanding of how to construct
symbolic encodings in two limiting cases only, namely for hyperbolic
systems and non-invertible maps of an interval into
itself~\cite{Guckenheimer83book,Katok95a,Hao89a,Badii97a}.
Unfortunately, most experimental low-dimensional systems fall outside
these two categories, except when they are sufficiently dissipative so
that their return maps can be modeled by one-dimensional maps.

To generalize one-dimensional symbolic dynamics to two-dimensional
invertible maps and hence to flows, methods have been proposed that
proceed by localizing homoclinic tangencies, i.e., points where the
stable and unstable manifold of the attractor are tangent to each
other~\cite{Grassberger85a}. Because this involves computing tangent
maps and estimating their eigendirections, these methods require that
the evolution equations are known, or at least that a model of the
dynamics is available.

In this article, we present in detail a completely different approach.
It is based on a topological analysis of chaotic
data~\cite{Gilmore98a,Tufillaro92a,Solari96a,Mindlin90a,Mindlin91a},
and extracts information not only in the neighborhood of the
singularities, but from the geometrical structure of the whole phase
space. More precisely, the way in which stretching and folding act on
the infinite number of unstable periodic orbits (UPO) embedded in any
strange attractor is exactly reflected in the way these orbits are
knotted and intertwined.

Stretching and folding are intimately related to symbolic dynamics.
Because a systematic study of the knots and links realized by periodic
orbits is made possible by template theory~\cite{Birman83a,Ghrist97a}
and template analysis~\cite{Mindlin90a,Mindlin91a}, \emph{precise
  information about the symbolic dynamics of the UPO can be extracted
  from their topological invariants}. As we show in this work, this
information, combined with the knowledge of the locations of the
periodic points in the section plane, allows one to determine an
excellent approximation to the border of a generating partition. This
method does not involve the differentiable structure of return maps at
all, and uses the concept of distance only to define neighborhoods,
more precisely to determine which member of a set of reference points
is nearest to a given point.

As this approach has already been applied to experimental time series
from a modulated laser using a preliminary version of the algorithm
described here~\cite{Lefranc94a}, the primary goal of this article is
to provide numerical evidence of the validity of the method. We thus
apply it to more than 1500 UPO extracted from numerical simulations,
and show that it is possible to obtain partitions which have a simple
structure, yet are completely consistent with the topological
organization of the UPO: the set of symbolic names assigned by the
partition to the UPO corresponds to a set of orbits of the horseshoe
template which have exactly the same topological invariants as the
extracted ones. Direct evidence of the fact that partitions obtained
in this way are generating will be presented in the second part of
this work~\cite{PlumecoqP99b}.

The article is organized as follows. In the remaining of this
introduction, we recall the links between the geometric properties of
chaos (stretching and folding) and symbolic dynamics. We then briefly
review the approach based on homoclinic tangencies, and we finally
illustrate the connection between symbolic dynamics and knot theory.

This connection can be precisely stated using template
theory~\cite{Birman83a,Ghrist97a} and template
analysis~\cite{Mindlin90a,Mindlin91a}. Since this approach to chaotic
dynamics is not widely known, Sec.~\ref{sec:template} is devoted to a
review of its main concepts. We put emphasis on the relation between
the symbolic name of an orbit and its topological invariants by giving
examples of the analytical formulas linking them, and specify our
fundamental assumptions.

In Sec.~\ref{sec:partition}, we describe our algorithm in detail by
progressively building a generating partition for a sample set of UPO
extracted from numerical simulations of a modulated laser. We finally
obtain a partition that is localized with a precision of the order of
0.01\% of the attractor width. Last, we conclude by discussing
possible extensions and applications of our method.

\subsection{Stretching, folding, and symbolic dynamics}
\label{sec:symdyn}

A striking feature of nonlinear dynamical systems is that they can
display complex behavior even when obeying simple equations of motion.
This seemingly paradoxical fact can only be understood by using a
geometric description of the dynamical laws, in which they are
represented as transformations of a phase space into itself. As is by
now commonly known, there are simple such transformations that
generate chaotic behavior by combining stretching and folding
mechanisms (as in, e.g., the R\"ossler system).

In the last decades, several methods have been proposed to
characterize a strange attractor, and thereby the underlying
dynamics~\cite{Abarbanel93a}.  Not surprisingly, some of the most
popular measures of chaos are deeply linked with the existence of the
stretching and folding mechanisms.  For example, Lyapunov exponents
quantify the efficiency of stretching by estimating the rate of
divergence of infinitely close trajectories.  Spectra of fractal
dimensions, and especially the correlation dimension as computed with
the Grassberger-Procaccia algorithm, have been widely used to analyze
the fractal structure that results from the repeated action of
stretching and folding.

\begin{figure}[htbp]
  \begin{center}
    \leavevmode
    \includegraphics[height=5cm]{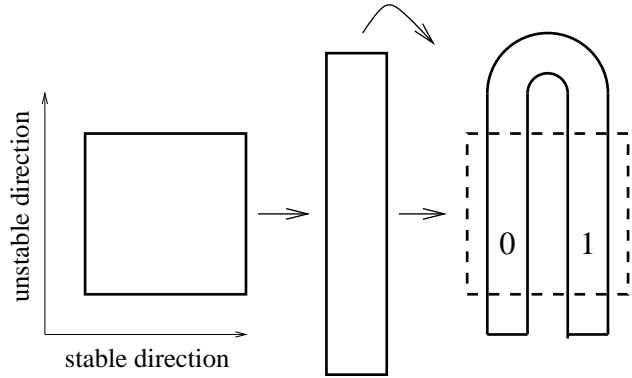}
    \caption{Representation of the action of the Smale's horsehoe map.
      A unit square is first stretched along the unstable direction
      and squeezed along the stable direction, then folded over itself
      so that it intersects the original square in two disjoint
      strips. }
    \label{fig:horseshoe}
  \end{center}
\end{figure}

Symbolic dynamics is another approach to chaotic dynamics that is
deeply rooted in the existence of the stretching and folding
mechanisms. The connection between symbolic dynamics and the geometric
properties of chaos is probably best illustrated by the paradigmatic
Smale's horseshoe map (Fig.~\ref{fig:horseshoe}), which is a key
example to understand the link between the geometric features of
chaos, symbolic dynamics and topological concepts.

If two points are not located on the same segment of the stable
manifold (i.e., along an horizontal line), their forward iterates will
eventually fall in different strips because of stretching. In the
opposite case, so will do the backward iterates because of squeezing.
Assigning distinct symbols to the two strips thus allows one to carry
out a symbolic dynamical study of this map, each point of the
invariant set being associated to a unique bi-infinite binary
sequence.  For a detailed presentation of the Smale's horseshoe map in
the context of topological analysis, see
Refs.\cite{Holmes86a,Holmes88a,Gilmore98a,Tufillaro92a,Solari96a}.

\begin{figure}[htbp]
  \begin{center}
    \leavevmode \includegraphics[width=10cm]{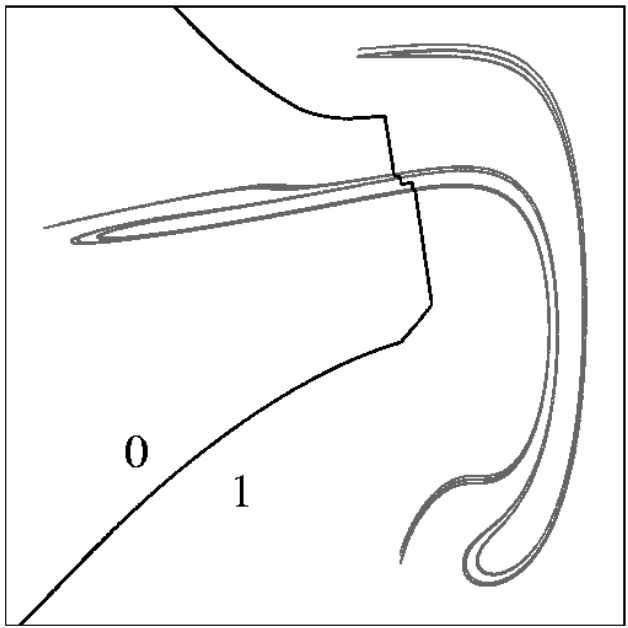}
    \caption{Symbolic encoding of a chaotic attractor using a
      partition of a section plane into two disjoint regions. This
      Poincaré section has been obtained from the modulated laser
      equations (\ref{eq:laser_eq}) described in
      Sec.~\ref{sec:partintro}.}
    \label{fig:coding}
  \end{center}
\end{figure}

In general, the symbolic encoding of a chaotic attractor is performed
by dividing a Poincar\'e section into a few disjoint regions
associated with distinct symbols (see Fig.~\ref{fig:coding}). In the
case of reversible equations of motion, each point of the attractor is
then associated to the bi-infinite sequence made of the symbols
corresponding to the regions visited by its backward and forward
iterates.

More precisely, consider a partition $\pa$ of the section plane $P$
in $n$ disjoint regions $\region_i(\pa)$, $i=0,..,n-1$. Assume that
for each point $x\in P$, $s_{\pa}(x)$ indicates the region which
contains $x$: $s_{\pa}(x)=i$ if $x\in\region_i(\pa)$.

The point $x$ is then represented by a bi-infinite symbolic sequence
\begin{displaymath}
\Psi(x)=
\{\ldots,\Psi_{-1}(x),\Psi_0(x),\Psi_1(x), \ldots\}, \quad
\Psi_i(x)=s_{\pa}(f^i(x)),
\end{displaymath}
where $f$ is the Poincar\'e return map.
Defining the shift operator $\sigma$ so that the sequence
$\Psi'(x)=\sigma\Psi(x)$ is made of the symbols
$\Psi'_i(x)=\Psi_{i+1}(x)$, it is readily seen that $\sigma$
represents the action of the return map in the space of symbolic
sequences, as $\sigma\Psi(x)=\Psi(f(x))$ by definition. 

Under certain conditions, such a coarse-grained measurement suffices
to provide an accurate description of the dynamics: two different
points, however close they may be, are associated to different symbol
sequences; the partition is then said to be
generating~\cite{Katok95a}. Of course, this is due to the amplifying
action of stretching, which connects arbitrarily small length scales
with large ones.

Symbolic dynamics can be given a rigorous foundation in the case of
hyperbolic systems, such as the Smale's horseshoe map shown in
Fig.~\ref{fig:horseshoe}. Indeed, hyperbolicity allows one to define
partitions (Markov partitions) that can be shown to be
generating~\cite{Katok95a}.  In this context, symbolic dynamics is of
utmost importance to prove several fundamental theorems of dynamical
system theory. For example, a symbolic dynamical analysis of the
horseshoe dynamics easily shows that the invariant set contains
aperiodic orbits, a dense infinity of unstable periodic orbits, and
that there is at least one orbit which is dense in the invariant
set~\cite{Guckenheimer83book,Katok95a}.

For non-hyperbolic systems, rigorous results are known only in the
case of non-invertible maps of an interval into itself, such as the
well-known logistic map. In this case, a generating partition is
obtained by dividing the one-dimensional interval into regions where
the map is monotonic: the border of the partition consists of the
critical points of the map, where the derivative
vanishes~\cite{Guckenheimer83book,Hao89a,Badii97a}.

However, most strange attractors encountered in experimental systems
or numerical simulations are non-hyperbolic: orbits are created and
destroyed as a control parameter is varied, which is incompatible with
the structural stability implied by hyperbolicity. Moreover,
one-dimensional symbolic dynamics can only be used for extremely
dissipative systems, and even then only in an approximate way. Whether
symbolic dynamics can be put on a sound basis in the general case thus
remains an open and fascinating problem.

A guiding fact is that the parameter space of a dynamical system such
as, e.g., the H\'enon map $(x,y)\rightarrow (a-x^2+b y,x)$ contains
generally both the hyperbolic and one-dimensional limits. For a
sufficiently large value of the $a$ parameter, the H\'enon map has an
invariant hyperbolic repellor; it becomes equivalent to the
one-dimensional map $x\rightarrow a-x^2$ when the $b$ parameter goes
to zero.  Therefore, a general procedure for constructing a symbolic
encoding of a non-hyperbolic, weakly dissipative, attractor should
have the one-dimensional and hyperbolic codings as limiting cases.

\subsection{Symbolic encodings based on homoclinic tangencies}
\label{sec:homtang}

Accordingly, the method proposed by Grassberger and
coworkers~\cite{Grassberger85a,Grassberger89a} is a generalization of
the one-dimensional theory. For a 1D map, the border of the partition
naturally consists of the critical points of the map, whose existence
is responsible for the non-invertibility of the map. In the case of
invertible 2D maps, there are no critical points, but their natural
counterparts are the \emph{homoclinic tangencies}, where the stable
and unstable manifolds of the attractor are tangent to each other.
Their existence stems from the non-hyperbolicity of the map: in a
sense, an invertible 2D map loses invertibility at homoclinic
tangencies when iterated an infinite number of times. Furthermore,
points of homoclinic tangency converge to backward and forward images
of the critical points of the 1D map when dissipation is increased to
infinity.

Grassberger and Kantz thus conjectured that a good symbolic encoding
could be obtained by dividing the plane with a line connecting
homoclinic tangencies \cite{Grassberger85a,Grassberger89a}. Several
studies have given numerical evidence that the partitions so obtained
were generating to a high level of accuracy
\cite{Grassberger85a,Grassberger89a,Cvitanovic88a,%
  DAlessandro90a,Giovannini91a,Jaeger97b}. Another motivation for this
rule is the fact that points located on opposite sides of a homoclinic
tangency converge to each other both for positive and negative time.
Thus, they can only be distinguished if they are associated to
different symbols.

However, this approach has been rarely used, if ever, to characterize
the symbolic dynamics of experimental chaotic time series (see however
Ref.~\cite{Wu96a} for an application to time series generated from
numerical simulations). Indeed, it heavily relies on the knowledge of
the equations of motion and on the computation of the tangent map to
determine the location of the homoclinic tangencies. While the
direction of the invariant manifolds could in principle be estimated
by fitting a model to the dynamics in the neighborhood of a
point~\cite{Wu96a}, the application of such a procedure to
experimental time series seems hazardous. Indeed, it is a known fact
that there is a dramatic noise amplification precisely at homoclinic
tangencies~\cite{Jaeger97a}: since the stable manifold is tangent to
the unstable manifold, it cannot drive perturbed trajectories back to
the attractor. In this situation, extracting information from a
tangent map constructed by estimating derivatives appears to be
problematic.

Furthermore, it should be noted that this method is faced with the
difficulty of choosing which homoclinic tangencies to connect, because
all images and preimages of a homoclinic tangency are themselves
homoclinic tangencies.  To address this problem,
Ref.~\cite{Giovannini91a} proposed to use only the so-called
``primary'' homoclinic tangencies, i.e., tangencies such that the sum
of the curvatures of the stable and unstable manifolds is smaller than
for all their images and preimages.  Another approach to solving this
problem was presented in Ref.~\cite{Jaeger97b}, where the global
organization of the lines of homoclinic tangencies in the phase space
was studied.

This ambiguity is due to the fact that techniques based on homoclinic
tangencies focus on the singularities induced by folding in the limit
of infinite time.  However, it is known from singularity theory (see,
e.g., Ref.~\cite{Gilmore81book}) that singularities at a point
organize the structure of an extended neighborhood of this point.
Accordingly, there should be prints of the folding process in the
whole phase space.

Indeed, there is another approach to the construction of symbolic
encodings that focuses on the global organization of the strange
attractor: it is based on a topological analysis of its unstable
periodic orbits. That topological invariants of an unstable periodic
orbit provide key information about the associated symbolic dynamics
was, to our knowledge, first noted by Solari and
Gilmore~\cite{Solari88a}. A method to construct a generating partition
based on this idea was then outlined by Lefranc \emph{et
  al.}~\cite{Lefranc94a} and applied to experimental time series from
a modulated laser.  Note that the fact that a generating partition
assigns different names to different periodic orbits has also
independently been used to construct symbolic encodings in
Refs.~\cite{Flepp91a,Finardi92a,Badii94a}.

\subsection{From unstable periodic orbits and knot theory to symbolic
  dynamics} 
\label{sec:upoknot}

A strange attractor is not the only invariant set of a chaotic
dynamical system, as it typically has embedded in it an infinite
number of unstable periodic orbits (UPO). While these UPO, whose
existence is due to ergodicity of chaotic dynamics, are known since
the works of Poincar\'e, they have only been fully utilized to
characterize and control chaos in the last decade (see, e.g.,
Refs.~\cite{Auerbach87a,lathrop89:_charac,ott90:_contr,Mindlin91a,Badii94a}).
As we see in the following, they also prove to be invaluable for
extracting symbolic dynamical information from experimental data.

\begin{figure}[htbp]
  \begin{center}
    \leavevmode
    \includegraphics[angle=270,width=10cm]{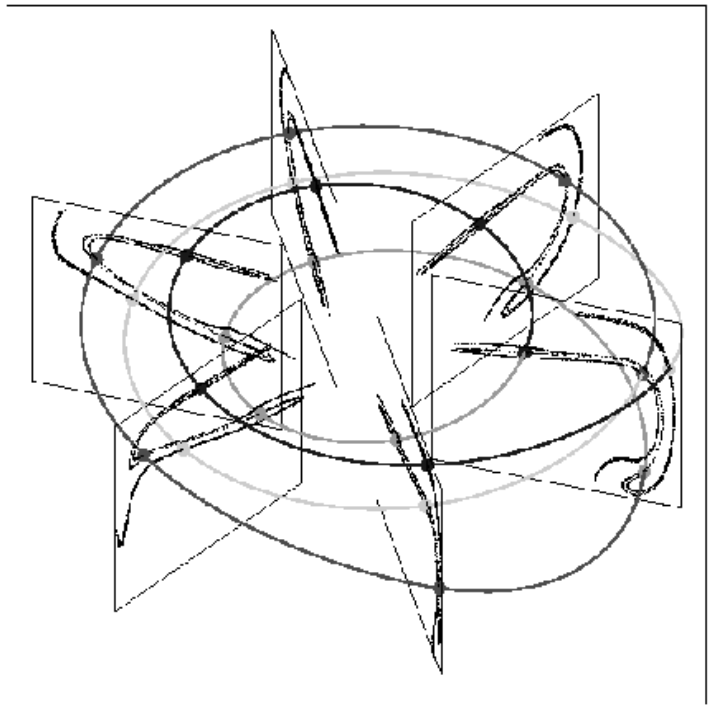}
    \caption{Stretching and folding braid a periodic orbit in a
      definite way, as can be seen here with a period-4 orbit. }
    \label{fig:braid4t}
  \end{center}
\end{figure}

As every trajectory in the attractor, unstable periodic orbits
experience stretching and folding. But, as they exactly return to
their initial condition in a short amount of time, they bear the mark
of these mechanisms in a very distinct way: their associated closed
curves in phase space are braided in a way that precisely reflects the
action of stretching and folding (see Fig.~\ref{fig:braid4t}). Because
symbolic dynamics is also intimately related to stretching and
folding, the way in which periodic orbits are intertwined must carry
symbolic dynamical information.

What makes this simple observation so fruitful is that this relation
can be expressed in well-defined mathematical terms for strange
attractors that can be embedded in a three-dimensional phase space.
Indeed, characterizing the topological structure of closed curves in
such a space is nothing but the central problem of knot theory (see
e.g.~\cite{Kaufmann91a}). Knot theory provides us with topological
invariants that can be utilized to decide whether two closed curves
can be continuously deformed into each other, i.e.  have identical
knot types or not, and thus to classify periodic orbits according to
their geometrical structure.

The relevance of knot theory in the context of dynamical system theory
stems from one of its fundamental theorems. Indeed, the uniqueness
theorem states that one and only one trajectory passes through a non
singular point of phase space (because of determinism). In particular,
this implies that a periodic orbit cannot intersect itself, nor
another orbit, and thus that the knots and links they form have a
well-defined type. Moreover, changing a control parameter will usually
change the shape of a periodic orbit but, for the same reason, will
not induce intersections.  Consequently, the knot type of a periodic
orbit remains unchanged on the whole domain of existence of the orbit,
and can be viewed as a genuine fingerprint.

It is thus obvious that topological invariants from knot theory
provide us with a robust way to characterize how stretching and
folding intertwine unstable periodic orbits. As an example, the
simplest topological invariant, the linking number, indicates how many
times one orbit winds around another. What makes these invariants
relevant for experimental studies is their robustness. If two periodic
orbits are sufficiently separated, knot invariants can be reliably
determined even when only approximate trajectories, possibly
contaminated by noise, are available (as typically extracted from a
time series). Indeed, the possible perturbations then merely amount to
small deformations of the orbit and do not change the invariants.

It should be noted that because the topological approach relies on
knot theory, it can only be applied to flows and hence to
orientation-preserving two-dimensional return maps. Thus,
orientation-reversing 2D return maps, such as the H\'enon map at the
standard parameters $(a=1.4,b=0.3)$, fall outside its scope. However,
this will allow us to show that phenomena that have been observed in
such maps~\cite{Hansen92a,Giovannini92a} violate the more restrictive
constraints obeyed by orientation-preserving return maps.

The link between topological invariants and symbolic dynamics is
provided by the tools of template theory. Since the main concepts of
the latter are not widely known, we review them in the next section,
before presenting the details of our algorithm in
Sec.~\ref{sec:partition}.

\section{Periodic orbits, knots and templates}
\label{sec:template}

\subsection{Template theory of hyperbolic systems: the
  Birman--Williams theorem} 
\label{sec:temtheo}

As is the case for many features of chaotic behavior, most of the
rigorous results about the topological structure of unstable periodic
orbits are known in the context of hyperbolic dynamical systems. They
compose what may be called template theory~\cite{Ghrist97a}. The
keystone of the latter is the Birman-Williams
theorem~\cite{Birman83a,Birman83b}, which shows that the topological
organization of the unstable periodic orbits of an hyperbolic flow can
be studied in a systematic way.

Given a hyperbolic chaotic three-dimensional flow $\Phi_t$ with an
invariant set $\Lambda$, let us define an equivalence relation between
points of $\Lambda$ in the following way:
\begin{equation}
  \label{eq:bw_equi}
  \forall x,y\in \Lambda,\quad x \sim y \Leftrightarrow
  \lim_{t\rightarrow\infty}||\Phi_t(x)-\Phi_t(y)||=0,
\end{equation}
which relates points having the same asymptotic future. Identifying
points in the same equivalence class thus amounts to collapsing the
invariant set along its stable manifold. The Birman-Williams
theorem~\cite{Birman83a,Birman83b}
consists of two main statements:
\begin{enumerate}
\item In the set of equivalence classes of relation
  \eqref{eq:bw_equi}, the hyperbolic flow $\Phi_t$ induces a semi-flow
  $\bar{\Phi}_t$ on a branched manifold $\mathcal{K}$. The pair
  $(\bar{\Phi}_t,\mathcal{K})$ is called a \emph{template}, or
  \emph{knot-holder}, for a reason that is made obvious by the second
  statement.
\item Unstable periodic orbits of $\Phi_t$ in $\Lambda$ are in
  one-to-one correspondence with unstable periodic orbits of
  $\bar{\Phi}_t$ in $\mathcal{K}$. Moreover, each unstable periodic
  orbit of $(\Phi_t,\Lambda)$ is isotopic to the corresponding orbit
  of $(\bar{\Phi}_t,\mathcal{K})$, the same property holding for any
  link made of a finite number of UPO. Thus, periodic orbits in the
  invariant set can be continuously deformed without any crossing so
  as to be laid on the branched manifold.
\end{enumerate}

The second statement implies that any topological invariant defined in
the framework of knot theory will take identical values on a set of
UPO of the flow and on the corresponding set of periodic orbits of the
template.

The proof of the Birman-Williams theorem relies on a key property: two
points belonging to the same periodic orbit, or to different periodic
orbits, have by definition different asymptotic futures; if initially
separated, they will remain at a finite distance forever. Thus, a
periodic orbit cannot intersect its own stable manifold, nor the
stable manifold of another orbit. As a result, \emph{collapsing the
invariant set along its stable manifold does not induce crossings
between periodic orbits, hence does not modify their topological
organization}.

This simple observation is central to template theory and template
analysis because it clearly shows that their concepts are insensitive
to the degree of dissipation, which becomes irrelevant after reduction
of the stable manifold. \emph{In a given topological class, any flow
  has the same global topological organization as an infinitely
  dissipative flow}.  This is precisely what will allow us to use
template analysis as a bridge between one-dimensional and
two-dimensional symbolic dynamics.

\begin{figure}[htbp]
  \begin{center}
    \leavevmode
    \includegraphics[angle=270,width=12cm]{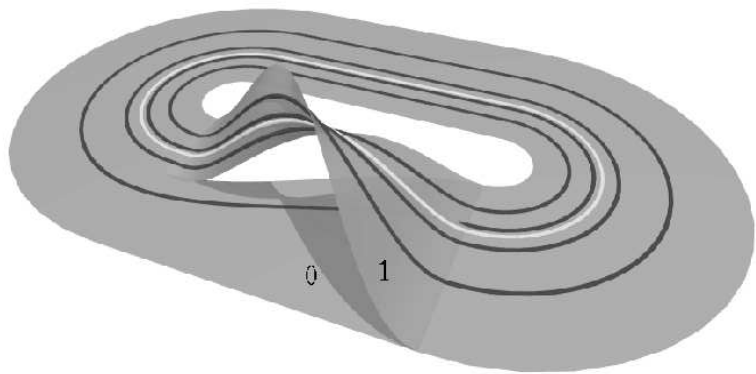}
    \caption{The Smale's horseshoe template, with a period-1 and a
      period-4 orbits. In this configuration, these orbit have exactly
      the same invariants they would have in a hyperbolic flow whose
      topological structure is described by this template. Because the
      branches correspond to the disjoint strips shown in
      Fig.~\ref{fig:horseshoe}, a symbolic description of the closed
      curves of the template can be given.}
    \label{fig:hstem}
  \end{center}
\end{figure}

As an example, the Smale's horseshoe template\footnote{by a slight
  abuse, the term ``template'' is often used to refer to the branched
  manifold alone, by assuming a standard structure for the semi-flow
  on the manifold.}, i.e. the branched manifold corresponding to a
flow whose return map is the Smale's horseshoe map, is shown in
Fig.~\ref{fig:hstem}. The number of branches, the torsions and linking
numbers of its branches define the structure of such a manifold, as
well as the order in which branches are stacked when they rejoin. The
Smale's horseshoe template presented in this form is an example of a
fully expansive template: the branches are stretched to the full width
of the template. This stretching, and the folding of branches over
each other describe geometrically the basic mechanisms of chaotic
dynamics. As will be recalled in Sec.~\ref{sec:symtem}, the
topological structure of a template can be concisely described by a
small set of integers which suffice to determine topological
invariants of a closed curve on the template, given its itinerary on
the branched manifold (i.e., the order in which it visits the
different branches).

For a more detailed exposition of the template theory of hyperbolic
sets, we refer the reader to
Refs.~\cite{Birman83a,Birman83b,Holmes85a,Holmes86a,Holmes87a,%
Holmes88a,Holmes89a}, and to a recent book by
Ghrist, Holmes and Sullivan~\cite{Ghrist97a} for a comprehensive
review.

\subsection{Template analysis of experimental systems}
\label{sec:temana}

The central problem of template theory is: given a hyperbolic
template, what can we say about the properties of knots living on this
template?

When we study an experimental system, however, the underlying template
is not a priori known, but unstable periodic orbits can be extracted
from time series, and their topological invariants and knot types
determined in a reconstructed phase space. Note that, while a strange
attractor is generally not hyperbolic, tools from template theory are
still relevant because the existing orbits should have the same
organization and the same invariants as in the hyperbolic limit,
provided they can be brought to this limit by a change in control
parameters.

In this context, the natural question then is: given a finite set of
knots contained in the attractor, can we construct a template which
holds all of them, and thus describes the global topological
organization of the strange attractor?

This program was pioneered by Mindlin \emph{et al.}~\cite{Mindlin90a},
who proposed to use the concepts of template theory to characterize
non-hyperbolic strange attractors by a small set of integers. They
demonstrated and thoroughly discussed the relevance of this approach
by showing in a beautiful work that all the topological invariants of
periodic orbits detected in time series from the Belousov-Zhabotinskii
chemical reaction allowed them to be globally laid on a Smale's
horseshoe template~\cite{Mindlin91a}. 

In the last decade, further evidence that the topological organization
of experimental chaotic systems could be described by templates has
been given in a variety of systems: a NMR
oscillator~\cite{Tufillaro91a}, CO$_2$ lasers with a saturable
absorber~\cite{Papoff92a,Fioretti93a}, or with modulated
losses~\cite{Lefranc93a,Lefranc94a}, a glow discharge~\cite{Braun95a},
a copper electro-dissolution reaction ~\cite{Letellier95b}, a
vibrating string~\cite{Tufillaro95a}, an electronic
circuit~\cite{Letellier96a}, a fiber laser~\cite{Boulant97b}, or a YAG
laser~\cite{Boulant97a}.  Similar conclusions have also been obtained
in numerical simulations of the Duffing~\cite{McCallum93a,Gilmore95a},
Lorenz~\cite{Letellier94a}, and R\"ossler
equations~\cite{Letellier95a}, and for systems modeling a bouncing
ball~\cite{Tufillaro94a}, pulsating stars~\cite{Letellier96c}, and
lasers~\cite{Boulant98a,Gilmore97a}.

All these studies follow more or less the same
procedure~\cite{Gilmore98a}. First, segments of time series shadowing
unstable periodic orbits are extracted from the experimental data, and
are embedded in a reconstructed phase space, where the topological
invariants of the associated closed curves are computed. Then the
simplest template on which the experimental orbits can be projected is
determined from the measured invariants. This is made possible by the
fact that the relevant information is carried by low-period orbits.
Indeed, the characteristic numbers of a template are completely
determined by the invariants of its spectrum of period-1 and period-2
orbits~\cite{Gilmore98a}.

The validity of a candidate template (determined from the
lowest-period orbits) can then be checked by verifying that the
invariants of the higher-period orbits allow them to be also laid on
the template. This is because the template characteristic numbers are
over-determined by the topological invariants of the unstable periodic
orbits. In the case of the Smale's horseshoe template, for example,
four integers suffice to compute the invariants of an infinite number
of periodic orbits.  As we will show in the following, the seemingly
redundant information carried by the topological invariants of a large
set of UPO can be used to extract information about the symbolic
dynamics of the attractor.

For further information, detailed introductions to template analysis
can be found in in a comprehensive review article by
Gilmore~\cite{Gilmore98a} and in books by Tufillaro, Abbott, and
Reilly~\cite{Tufillaro92a}, and by Solari, Natiello, and
Mindlin~\cite{Solari96a}.

\subsection{Extracting symbolic dynamical information from knot
  invariants}
\label{sec:symtem}

Our approach to the construction of symbolic encodings relies heavily
on the mathematical link between the topological invariants of
unstable periodic orbits and symbolic dynamics. To illustrate this
link more precisely, we now review briefly some of the basic tools of
template analysis.

As an example, we first consider the Smale's horseshoe period-$4$
orbit that is created in the initial period-doubling cascade.  We show
how its simplest invariant, namely its self-linking number\footnote{In
  the context of template analysis, the self-linking number is usually
  defined as the signed number of crossings of the braid representing
  the orbit.}, is easily computed from its symbolic name, which is
``0111'' if we use the coding shown in Fig.~\ref{fig:hstem}.

The branch line of a fully expansive template (the line where the
different branches are squeezed over each other) is a one-dimensional
analogue of a global Poincar\'e section: each period-$n$ orbit
intersects the branch line in exactly $n$ points. Because template
orbits cannot intersect on the two-dimensional (branched) manifold,
the layout of a periodic orbit on a template is completely determined
by the order in which its intersections with the branch line are
visited.

Computing this order is a classic exercise in symbolic dynamics of
maps of an interval into itself~\cite{Guckenheimer83book,Hao89a} (see
Refs.~\cite{Gilmore98a,Tufillaro92a,Holmes86a} in the context of
template analysis), since the return map of the branch line is
one-dimensional.  In the case of the Smale's horseshoe template, this
return map has the same structure as the standard logistic map, with
the region of positive (resp. negative) slope corresponding to the
branch with a torsion of zero (resp. one) half-turns. In our example,
it is easily found that the periodic points of the period-$4$ orbit
are found on the branch line in the order: ``0111'', ``1101'',
``1110'', ``1011''.

\begin{figure}[htbp]
  \begin{center}
    \leavevmode
    \includegraphics[width=10cm]{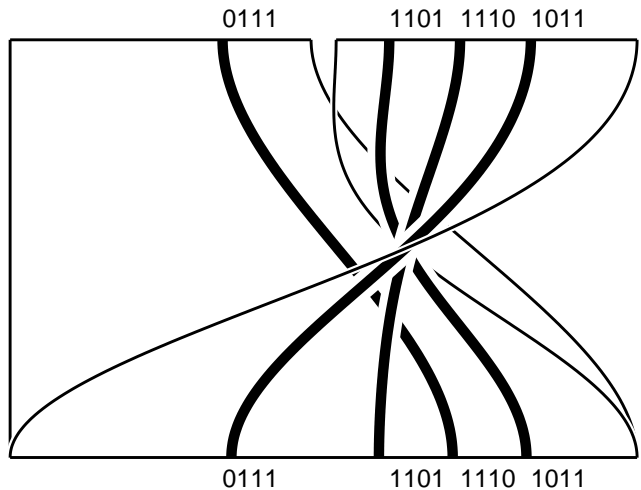}
    \caption{Geometry of the period-$4$ orbit ``0111'' on the
      template. Only the nontrivial part (i.e., the branched part) of
      the manifold is shown, as the top line can be identified with
      the bottom line (the branch line).  The layout of the periodic
      points on the branch line is completely fixed by the fact that
      branches ``0'' and ``1'' are orientation-preserving and
      orientation-reversing, respectively. Given the geometrical
      structure of the branches, this layout determines in turn the
      braid associated with the orbit, and hence all its topological
      invariants.}
    \label{fig:template_4t}
  \end{center}
\end{figure}

As can be seen in Fig.~\ref{fig:template_4t}, it then suffices to
connect periodic points to their images by following the semi-flow on
the branched manifold to obtain the braid associated with the orbit.
In this case, it is straightforward to verify that the self-linking
number of the ``0111'' orbit of the Smale's horseshoe template is 5.

This simple example illustrates concisely the key idea that \emph{the
symbolic dynamics of an unstable periodic orbit completely determines
its knot invariants and that conversely, the latter carry important
information about the former}. We now want to stress that this property
can be expressed by simple algebraic relations.

Following Mindlin \emph{et al.}~\cite{Mindlin90a,Mindlin91a}, the
structure of a $n$-branch template can be algebraically described by a
$n\times n$ matrix, the \emph{template matrix}, and a $1\times n$
matrix, the \emph{layering matrix}.  The template and layering
matrices are related to invariants of low-period orbits in the
following way.

Because the semi-flow on the branch manifold is expanding, each branch
carries one and only one period-$1$ orbit. The template matrix $t$ is
obtained from the organization of these period-$1$ orbits as follows.
The diagonal elements $t_{ii}$ indicate the local torsion of the orbit
on branch $i$, i.e. the rotation of its stable and unstable manifolds
in units of $\pi$. Off-diagonal elements $t_{ij}=t_{ji}$ are equal to
twice the linking number of the orbits located on branches $i$ and
$j$. In the case of the Smale's horseshoe with zero global torsion
shown in Fig.~\ref{fig:hstem}, with branches labeled ``0'' and ``1'',
the template matrix reads:

\begin{equation}
  \label{eq:hs_tem_matrix}
  t^{HS} =
  \left(
    \begin{array}[c]{cc}
      0 & 0\\
      0 & 1
    \end{array}
  \right)
\end{equation}
where $t_{11} = 1$ describes the folding of the ``1'' branch.

To complete the description of the template structure, one has to
specify in which order the different branches are superimposed when
they are glued together. Mindlin \emph{et al.}  define the $1\times n$
layering matrix $l$, which verifies $l_i<l_j$ iff branch $j$ is
located above branch $i$ on the branch line. Since the twisted branch
of the horseshoe template is folded over the untwisted one, its
layering matrix is given by:
\begin{equation}
  \label{eq:hs_layering}
  {l}^{HS} = \left(\,0\, 1\, \right)
\end{equation}

We use a slightly different convention and introduce a $n\times n$
symmetric matrix $l'$ such that for $i<j$, $l'_{ij} = 1$ if branch $j$
is located above branch $i$ and $l'_{ij} = -1$ otherwise (i.e.,
$l'_{ij}=-1$ indicates that the order of two branches differs from
that of a standard layering graph as defined in
Ref.~\cite{Melvin91a}). For the horseshoe template, we thus have:
\begin{equation}
  \label{eq:hs_layering2}
  {l'}^{HS} = \left(
    \begin{array}[c]{cc}
      0&1\\
      1&0
    \end{array}
    \right)
\end{equation}

A key property of template analysis is that simple analytic formulas
can be written to express some of the topological invariants of the
UPO as a function of the elements of the template and layering
matrices~\cite{Lefranc98pre_a}, using techniques similar to these
described in appendix E of Ref.~\cite{Tufillaro92a}. These invariants
are the (self-) linking numbers, relative rotation
rates~\cite{Solari88a}, and torsions of the periodic orbits. To
predict more sophisticated invariants, such as knot polynomials, a
description of the template as a framed braid (see, e.g.,
Ref.~\cite{Melvin91a}) would be required.

For example, the self-linking number of the ``0111'' orbit is given
for a general template by:
\begin{equation}
  \label{eq:slk_0111}
  slk(0111) = 3 t_{01} +3t_{11}+(3-\pi(t_{11}))\times l'_{01}
\end{equation}
where $\pi(t) = 1 (0)$ if $t$ is odd (even).  The reader may verify
that the value of 5 that can be obtained from
Fig.~\ref{fig:template_4t} is recovered by inserting in
Eq.~(\ref{eq:slk_0111}) the horseshoe template matrices given in
Eqs.~(\ref{eq:hs_tem_matrix}) and (\ref{eq:hs_layering2}).  Similar
expressions can easily be obtained for invariants of orbits of
arbitrarily high period. For example, we have:
\begin{eqnarray}
  \label{eq:lk_example}
  lk(01^{3} 0101^{2}, \left(01^{2}\right)^{3} 01^{4} 01^{3})&=&
  \frac{15}{2} t_{00}+39 t_{11}+\frac{69}{2}t_{01}\\
  &&+\left(
    30-\frac{19}{2} \pi(t_{11})\nonumber 
    -\frac{1}{2} \pi(t_{00}+t_{11})
    \right)\times l'_{01}
\end{eqnarray}
where $lk(\alpha,\beta)$ denotes the linking number of orbits $\alpha$
and $\beta$.

A crucial property of these expressions is that, except for the
presence of the terms involving the $\pi$ function, they are linear in
the elements of the template matrices $t$ and $l'$.  This is what
allows one to design a powerful algorithm to determine these elements
from the topological invariants of a few orbits of low period: one
considers all the possible symbolic names for these low-period orbits,
and all the possible branch parities, and selects those that lead to a
consistent, over-determined, set of linear equations. The solution to
such a set of equations is a candidate template, whose validity has
then to be checked with higher-period orbits.  The general procedure
will be described elsewhere~\cite{Lefranc98pre_a}, but some examples
may be found in Refs.~\cite{Boulant97b,Boulant98a}.

When the geometry of the branched manifold of the template has been
determined in this way, we then find all sets of symbolic names such
that template orbits with these names have exactly the same invariants
as the experimental periodic orbits. This indicates the different
possible projections of the set of UPO on the branched manifold that
preserve its topological organization.

In fact, there are only a few possible such projections for a given
experimental orbit. For example, in the case of the Smale's horseshoe
template, there is one and only one period-$7$ orbit of even torsion
with a self-linking number of 16: this is the ``0101011'' orbit. In
this case, the symbolic name of this orbit can be unambiguously
extracted from its topological structure.  In some other cases, there
may be several possible symbolic names. For example, the horseshoe
orbits ``001101'' and ``001011'' correspond to isotopic knots and thus
cannot be distinguished using the self-linking number or self-relative
rotation rates. However, they often can be identified using other
orbits which link them differently (if these orbits are found in the
attractor): in the previous example, there are four period-$8$
horseshoe orbits whose linking numbers with the two period-$6$ orbits
are different (e.g., $lk(00101011,001101) = 15$ but
$lk(00101011,001011) = 14$.)

\begin{table}[htbp]
  \caption{Basic topological properties of the periodic orbits with period
    up to 9 extracted from numerical simulations of a modulated
    laser model (see Sec.~\ref{sec:partintro}), and that will be used
    as an example in Sec.~\ref{sec:partition}. The listed invariants are:
    period, self-linking number, torsion. The symbolic names of
    horseshoe orbits with the same invariants are also displayed.
    Note that except for orbits 18 and 19, there is a single
    possible symbolic name. The two possible symbolic names for
    orbits 18 and 19 are related through a time-reversal symmetry of
    the Smale's horseshoe template.}
  \label{tab:symnames}
  \begin{tabular}[c]{|rrl|rrl|}
    \hline\hline
    Orbit & Invariants & Names &
    Orbit & Invariants & Names\\
    \hline
    1a & 1,0,1 & ``1''         & 8b & 8,21,5 & ``01011011''\\
    2a & 2,1,1 & ``01''        & 8c & 8,25,7 & ``01111111''\\
    4a & 4,5,3 & ``0111''      & 8d & 8,25,6 & ``01011111''\\
    5a & 5,8,3 & ``01011''     & 8e & 8,23,5 & ``01010111''\\
    5b & 5,8,4 & ``01111''     & 9a & 9,28,7 & ``011011111''\\
    6a & 6,13,5 & ``011111''   & 9b & 9,28,6 & ``010110111'', ``010111011''\\
    6b & 6,13,4 & ``010111''   & 9c & 9,28,6 & ``010110111'', ``010111011''\\
    7a  & 7,16,5 & ``0110111'' & 9d & 9,28,5 & ``010101011''\\
    7b & 7,16,4 & ``0101011''  & 9e & 9,30,7 & ``011101111''\\
    7c & 7,18,6 & ``0111111''  & 9f & 9,32,8 & ``011111111''\\
    7d & 7,18,5 & ``0101111''  & 9g & 9,30,6 & ``010101111''\\
    8a & 8,21,6 & ``01101111'' & 9h & 9,32,7 & ``010111111''\\
    \hline\hline
  \end{tabular}
\end{table}

This important fact is illustrated by Table~\ref{tab:symnames} which
shows an example where the symbolic names of all orbits up to period 9
extracted from an attractor, except two of them, can be obtained using
only the simplest topological invariants. This implies that there are
only two sets of horseshoe orbits which reproduce the measured
invariants. These sets differ by the names given to orbits 9b and 9c.

Although it should be noted that it is more common for higher-order
orbits to have several possible symbolic names, it appears very
clearly from Table~\ref{tab:symnames} that topological invariants
carry a large amount of information on the symbolic dynamics of a
chaotic system. As we now explain in Sec.~\ref{sec:fund_hyp}, this is
the property which the following of the article will rely on.

\subsection{Topological encoding as a bridge between the one-dimensional and
  the hyperbolic encodings}
\label{sec:fund_hyp}

As mentioned in Sec.~\ref{sec:upoknot}, the topological structure of a
given unstable periodic orbit is not modified by a change in a control
parameter. If we assume that there is a parameter that allows us to
freely tune dissipation, modifying this parameter will induce isotopic
deformations of the unstable periodic orbits, thus preserving their
topological structure (except for orbits that are annihilated or
created in saddle-node bifurcations).

Returning to the example of Table~\ref{tab:symnames}, let us
appropriately vary this control parameter so as to achieve infinite
dissipation. In this limit, the dynamics should be modeled by a
one-dimensional return map similar to the logistic map
$x_{n+1}=a-x_n^2$. Because of the deep link between template theory of
the Smale horseshoe and the symbolic dynamics of the logistic map, it
is then obvious that the symbolic name given by one-dimensional
symbolic dynamics theory will coincide with the one singled out by
topological analysis and indicated in Table~\ref{tab:symnames}.

Let us now assume that by varying another parameter, we bring the
system to a region of parameter space where it has an hyperbolic
invariant set. Because template theory is mathematically rigorous in
this case, the topological symbolic names in Table~\ref{tab:symnames}
must also be consistent with the ones obtained from a canonical Markov
partition.

We will therefore make the fundamental hypothesis that \emph{any
  relevant symbolic encoding should assign to a given periodic orbit a
  name that is compatible with its topological structure}, i.e. such
that the orbit with the same name on the associated template has
identical topological invariants. This is a strong assumption, as it
implies that an orbit with a single topological name must be assigned
the same name on its whole domain of existence (provided the global
topological structure described by the template is not modified).
However, this appears to be the only way to connect the two limiting
cases in a continuous way.

Because this assumption is central to the method we describe below, it
is important to note that it might seem to be contradicted by some
observations reported in the literature. In particular, Hansen has
described the following striking phenomenon~\cite{Hansen92a}: by
following a certain closed loop in the parameter space of the H\'enon
map starting and ending at parameters $(a=2,b=0)$ (where the
one-dimensional canonical coding is available), the unstable periodic
orbit with initial symbolic name ``011111'' is transformed into the
orbit with symbolic name ``000111''. This observation seems to
indicate that there cannot be a global symbolic name in the whole
parameter space. Similarly, Giovannini and Politi have pointed out
that at some parameter values, the symbolic encodings of some periodic
orbits can experience sudden changes due to the annihilation of
primary homoclinic tangencies~\cite{Giovannini92a}.

Although the H\'enon map is generally considered to capture the essential
features of low-dimensional chaotic dynamics, we believe that it would
be incorrect to conclude from these studies that such discontinuities
occur in all two-dimensional maps involving the creation of a
horseshoe. More precisely, we highly suspect that the situation is
dramatically different for orientation-preserving maps (i.e., maps
that can be viewed as return maps of a three-dimensional flows), as
the simple following argument shows.

In any suspension of a horseshoe-type map with zero global torsion,
the self-linking number of the ``011111'' orbit is 13. There is only
one other period-$6$ horseshoe orbits corresponding to this value: the
``010111'' orbit which is the saddle-node partner of the ``011111''
orbit\footnote{It should be noted that orbits born in the same
  saddle-node bifurcation are always isotopic for an obvious reason.}.
Since the latter can be distinguished from the former in that it has
odd torsion (i.e., negative Floquet multipliers), there is absolutely
no way in which the unstable ``011111'' orbit could be turned into
another orbit by following a closed path in parameter space,
\emph{provided this path stays on the orientation-preserving side of
  parameter-space}. Under this condition, indeed, a suspension of the
H\'enon map that deforms continuously as the parameter is varied can
easily be constructed.

Similarly, Giovannini and Politi have reported that at some parameters
of the H\'enon map (also in the orientation-reversing case), the
partition line experiences a discontinuity in such a way that the
substring ``...11000...'' has to be replaced by the substring
``..01001...'' in all the symbolic names of the periodic
orbits~\cite{Giovannini92a}. Defining a $C_n$ orbit as an orbit whose
symbolic name contains the substring ``$0^{n-1}$'', but not
``$0^{n}$'', this recoding would turn some $C_4$ orbits into $C_3$
orbits. Although one can find some examples of $C_3$ and $C_4$ orbits
that have the same braid type, the two classes of orbits are easily
distinguished through their linking numbers with other
orbits~\cite{Lefranc94a,Solari88a}. Thus, we have here another
phenomenon where the change in the symbolic name is accompanied by a
change in the topological invariants associated with this name.

Of course, our argument does not prove that the Hansen phenomenon
cannot occur for higher-order orbits with a larger number of isotopic
orbits (if they cannot be distinguished through their linking numbers
with a third orbit). However, this clearly shows that there are
effects which occur in orientation-reversing maps which would violate
the uniqueness theorem in suspensions of orientation-preserving maps
(because these effects connect symbolic names corresponding to
different topological invariants).  This somehow questions the
relevance of the H\'enon map at the classical parameters (where the
Jacobian is negative) as a prototype of return maps in
three-dimensional flows.

There is thus, to our knowledge, currently no clear counter-example to
our hypothesis that well-defined symbolic encodings can be obtained
for orientation-preserving maps. Therefore, we now proceed and
describe how to construct generating partitions that are compatible
with the topological invariants of the UPO. We will see that, while
the algebraic tools of template analysis do not always select a single
name for every orbit, a non-ambiguous encoding and a complete
identification of the symbolic names are eventually obtained if we
additionally require the symbolic encoding to be continuous, so that
points which are close in a section plane are encoded by sequences
that are close in sequence space.

\section{Description of the algorithm}
\label{sec:partition}

\subsection{Detection of the unstable periodic orbits}
\label{sec:partintro}

As we have seen in Sec.~\ref{sec:template}, template analysis yields
for each detected periodic orbits a list of possible symbolic names.
The next step is to use this information and the locations of these
periodic orbits in the section plane to construct a partition, which
may then be used to encode chaotic trajectories as well.

To illustrate the procedure that we detail below, we will study a
chaotic attractor observed in numerical simulations of a modulated
class-B laser, described by the following
equations~\cite{Tredicce86a,Dangoisse87a}:

\begin{subequations}
  \label{eq:laser_eq}
\begin{eqnarray}
  \dot{I}&=&I [AD-1-m\sin\omega t]\\
  \dot{D} &=& \gamma\left[1-D (1+I)\right]
\end{eqnarray}
\end{subequations}
where the variables $I$ and $D$ represent the output intensity and the
population inversion. In our numerical simulations, the following
parameters were used: $A=1.1$ (pump rate), $m = 0.0334$ (modulation
amplitude), $T=2\pi/\omega=300$ (modulation period), and $\gamma =
2.5\times 10^5/1.2\times 10^8 =2.083\times 10^{-3}$ (ratio of the
population inversion relaxation rate to the cavity damping rate).
Fig.~\ref{fig:coding} shows the Poincar\'e section in $(\log I,D)$
coordinates corresponding to $t=0 \bmod T$ (as for all the Poincar\'e
sections shown in this paper). The topological structure of this
attractor is described by the Smale's Horseshoe template shown in
Fig.~\ref{fig:hstem}.

The algorithm we describe in this section will allow us to determine
unambiguously the symbolic names of a set of unstable periodic
embedded in the strange attractor. If we want to utilize this
information to perform symbolic encodings of arbitrary trajectories,
we must detect a set of orbits that provides a good cover of the
attractor, i.e., which is such that all trajectories on the attractor
are locally shadowed with a good precision by an UPO.

Our detection code was specially designed to achieve this goal.
Basically, it divides the Poincar\'e section in cells of size
$\epsilon$ and follows a long chaotic trajectory, searching for close
returns. When one is found, we check whether all the cells visited by
points in the corresponding time series segment contain periodic
points of period lower than or equal to the recurrence time. If this
is not the case, a Newton-Raphson iteration is started from this
initial condition. When the latter succeeds, the quality of the cover
has been improved. The search terminates when each cell contains at
least one periodic point and when no significant improvement has been
obtained over a certain interval of time (the detection of a periodic
point of lower period than those already contained in the cell is
considered as an improvement). In this way, the computational effort
is concentrated on obtaining the most uniform cover with orbits of
lowest periods, rather that finding the highest possible number of
orbits.

This preliminary investigation revealed an interesting property: some
parts of the strange attractor are extremely difficult to shadow with
orbits of low period, especially when there are a lot of forbidden
sequences in the symbolic dynamics. It turns out that these regions
will be found later to be close to the partition border and to lines
of homoclinic tangencies. If we view periodic and chaotic trajectories
as the analogues of rational and irrational numbers, respectively,
this observation could be rephrased as: near principal lines of
homoclinic tangencies, chaotic trajectories are more ``irrational''
than elsewhere in the attractor.  While this may seem to be a
fundamental obstacle to our approach, it should be noted that because
the dynamics is weakly unstable in these regions, it is easy to detect
the high-period orbits which are located in them, and that topological
invariants of high-period orbits can be computed robustly. This
explains why, in spite of the above-mentioned effect, we will be able
to localize partition borders to within 0.01\% of the attractor width
in Sec.~\ref{sec:optim}. Furthermore, we will show in the second part
of this work~\cite{PlumecoqP99b} that because of non-hyperbolicity,
obtaining a high-resolution shadowing in these regions is in fact not
at all crucial for characterizing accurately the symbolic dynamics.

Following the procedure described above with $\epsilon=0.001$ and with
a maximal period of $32$, we obtained a set of 1594 periodic orbits
providing a uniform cover of the attractor. This set of orbits will be
used throughout this section to illustrate the different stages of our
algorithm. The possible symbolic names of the lowest-period orbits as
determined from template analysis have been given in
Table~\ref{tab:symnames}.

\subsection{Notations}
\label{sec:notation}

The detected set of orbits will be noted $\mathcal{O}$, and consists
of $\orbnum$ UPO $O_i$. Each periodic orbit $O_i$ has $p_i$
intersections $O_i^j$, $j=1,\ldots, p_i$, with the section plane ($p_i$
is the topological period of the orbit). These intersections are
periodic points of the first return map $f$, and their set will be
noted $\mathcal{P}$.

As we have seen in Sec.~\ref{sec:template}, knot theory and template
analysis provide us for each orbit $O_i$ with one or several possible
names. These ``topological names'', which will be noted
$\nameofi{O_i}{k}$, are the names of the template orbits which have
the same topological invariants. For definiteness, and since all
cyclic permutations of a topological name represent the same orbit, we
always write the topological name using the lowest permutation in the
lexicographic order, enclosed inside brackets.  For example, if the
period-$2$ orbit $O_2$ can be named ``01'' or ``10'', then
$\nameof{O_2}=\orbit{01}$.

Symbolic names are also used to label periodic points. In this
case, cyclic permutations of a given string of symbols are not
equivalent, since they correspond to different periodic points. In
this context, we use overlined strings. For example, the intersection
of the orbit $O_2$ with the section plane consists of two periodic
points: $\perpoint{01}$ and $\perpoint{10}$.

A partition $\pa$ of the section plane into $n$ disjoint regions
$\region_i(\pa)$ assigns to each UPO a symbolic name
$\pnameof{O_i}{\pa}$. Because two partitions that associate a given
periodic point with different cyclic permutations of the same name are
to be considered different, we define $\pnameof{O_i}{\pa}$ as being
the symbolic name of its first periodic point:
$\pnameof{O_i}{\pa}$=$\pnameof{O_i^1}{\pa}$. The latter is made of the
symbols associated with the regions containing by $O_i^1$,
$O_i^2$,\ldots,$O_i^p$.

\subsection{Parameterization of partitions by periodic orbits}
\label{sec:part_paramn}

Let us first consider the period-$1$ and period-$2$ orbits $O_1$ and
$O_2$ whose symbolic names can be unambiguously determined as being
$\nameof{O_1}=\orbit{1}$ and $\nameof{O_2}=\orbit{01}$. The latter
consists of two periodic points whose symbolic names are the cyclic
permutations of $\nameof{O_2}$, namely $\perpoint{01}$ and
$\perpoint{10}$.

There are thus two possibilities for assigning a symbolic sequence to
the two points $O_2^1$ and $O_2^2$ of the $O_2$ orbit.  Either $\{
\pnameof{O_2^1}{\pa}, \pnameof{O_2^2}{\pa}\} =
\{\perpoint{01},\perpoint{10}\}$ or the opposite choice is made. As we
will see in the second part of this work~\cite{PlumecoqP99b}, these
two possibilities lead to different, but dynamically equivalent,
solutions. For definiteness, we restrict ourselves to the first
configuration in this section.

In the following, we call \emph{reference points} the periodic points
whose symbolic sequence is assumed to be unambiguously known. We now
explain how the three reference points $\{O_1^1,O_2^1,O_2^2\}$,
associated with sequences
$\{\perpoint{1},\perpoint{01},\perpoint{10}\}$, may be used to define
a rough partition, which will be later refined by considering
higher-order periodic orbits.

If we examine generating partitions such as these shown in
Fig.~\ref{fig:coding} and in
Refs.~\cite{Grassberger85a,Grassberger89a,DAlessandro90a,Giovannini91a},
we note that the regions of the section plane corresponding to
different symbols are separated by a line with a simple structure,
whose length is of the order of the diameter of the attractor.
Consequently, there is a high probability, as higher as the points are
closer, that a point and one of its close neighbors correspond to the
same symbol, except if they are located in a small region around the
border.

If a point is in a close neighborhood of one of the three reference
points, it is natural to encode this point with the same symbol as
this reference point. For points which are at comparable distances
from two or more reference points, the correct symbol is uncertain.
However, without using the information that will be provided by the
higher-order periodic orbits, the simplest procedure that is
consistent with the previous remark is to associate these points with
the symbol of the closest reference point. We thus have a simple rule
to encode a chaotic trajectory: at each intersection with the section
plane, the closest reference point is determined and the associated
symbol is inserted in the symbolic sequence.

\begin{figure}[htbp]
  \begin{center}
    \leavevmode
    \includegraphics[angle=270,width=10cm]{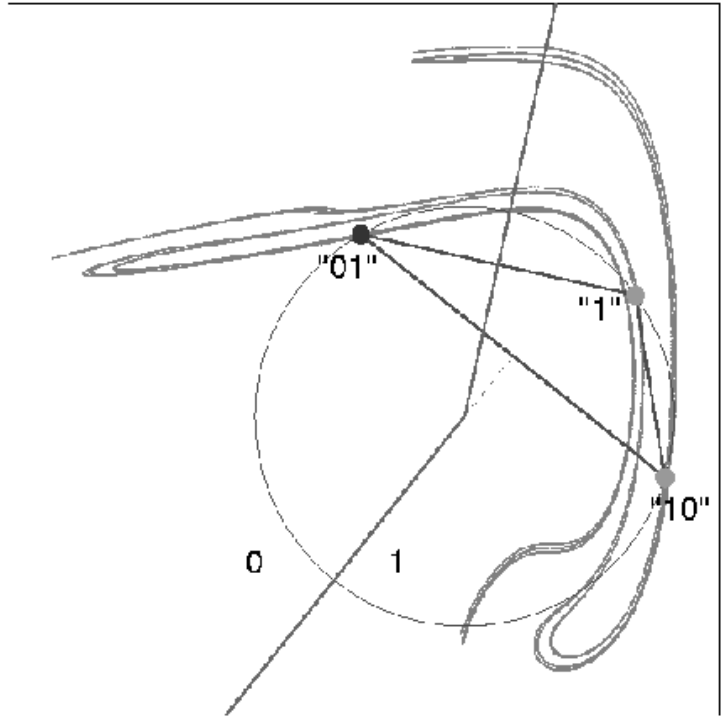}
    \caption{The initial partition based on periodic points
      $\perpoint{1}$,$\perpoint{10}$, and $\perpoint{01}$. For points
      that are at the left (resp. right) of the border, the closest
      reference point is the $\perpoint{01}$ periodic point (resp. one
      of the $\perpoint{1}$ and $\perpoint{10}$ periodic points). The
      circumcenter of the triangle made of the three points is also
      shown. }
    \label{fig:partinit}
  \end{center}
\end{figure}

The corresponding partition of the section plane obtained using the
three initial reference points is shown in Fig.~\ref{fig:partinit}. In
this simple case, the border line of the partition is easily
constructed, since one has merely to separate points whose nearest
reference point has leading symbol ``0'' from those whose nearest
reference point has leading symbol ``1''. Thus, the partition border
follows the mediators of the segments joining points with different
symbols (i.e., the segments from $\perpoint{01}$ to $\perpoint{1}$ and
from $\perpoint{01}$ to $\perpoint{10}$). It is a known geometrical
property that these two mediators intersect at the circumcenter of the
triangle made of the three initial reference points.

A nice property of the above rule is that it can be efficiently
implemented for an arbitrary number of reference points, using
well-known geometrical tools: Delaunay triangulations and Vorono\"\i{}
diagrams~\cite{aurenhammer91:_voron,preparata85:_comput_geomet,%
okabe92:_tessel}. 

\label{sec:partinit}
\begin{figure}[htbp]
  \begin{center}
    \leavevmode
    \includegraphics[angle=270,width=12cm]{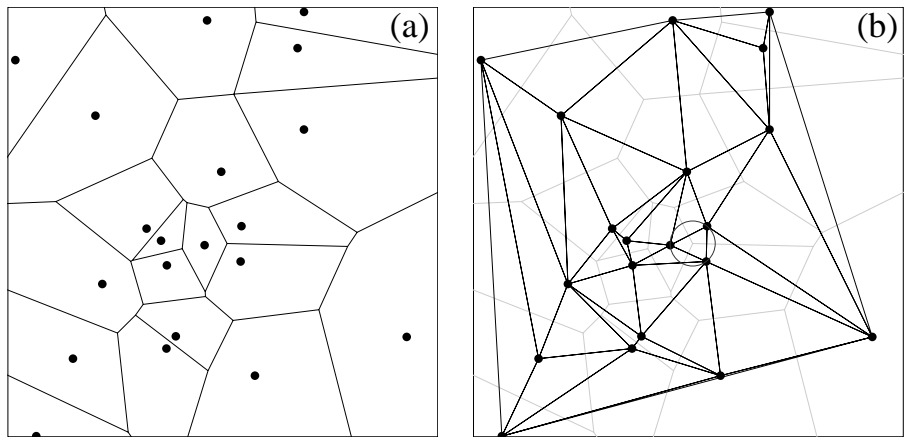}
    \caption{(a) Vorono\"\i\ diagram of a set of points. (b) the
    corresponding Delaunay triangulation. The circumcenter of one of
    the triangles is shown.}
    \label{fig:voronoi}
  \end{center}
\end{figure}

Given a reference point $O_i^j$, the set of points in the section
plane that are closer to $O_i^j$ than to any other reference point is
nothing but the Vorono\"\i\ domain of $O_i^j$ with respect to the set
of reference points. The Vorono\"\i\ diagram is a graph that consists
of the borders of the Vorono\"\i\ domains (Fig.~\ref{fig:voronoi}a).
The dual graph of the Vorono\"\i\ diagram is called the Delaunay
triangulation (Fig.~\ref{fig:voronoi}b). Among the possible
triangulations of a set of points, the Delaunay triangulation is the
only one such that the circumcircle of a triangle linking three sites
never contains another
site~\cite{aurenhammer91:_voron,preparata85:_comput_geomet,okabe92:_tessel}.
This property can be used to implement efficient algorithms for
building Delaunay triangulations, from which the associated
Vorono\"\i\ diagrams is easily obtained. Delaunay triangulations will
thus be a powerful tool to construct partitions and parameterize them
in a way that is suitable for applications.

In our initial configuration based on three reference points, the
Delaunay triangulation is readily obtained since it merely consists of
the triangle made of the three initial reference points
(Fig.~\ref{fig:partinit}). As explained above, the Vorono\"\i\ domains
of the three points are separated by the mediators of the triangle
edges, which intersect at the center of the circumcircle. The ``0''
(resp. ``1'') region consists of the Vorono\"\i{} domain of
$\perpoint{01}$ (resp.  the union of the Vorono\"\i{} domains of
$\perpoint{1}$ and $\perpoint{10}$).

To determine the border line for triangulations with an arbitrary of
reference points, one searches for couples of neighboring triangles
whose common edge carries two different symbols.  The line segments
connecting the circumcenters of all such pairs of triangles constitute
the border line. This allows one to compute quickly the partition
corresponding to a given set of reference points. Another advantage of
Delaunay triangulations is that they can be computed incrementally:
adding a new reference point to an existing triangulation only
requires modifying the triangles in the neighborhood of the new
point~\cite{watson81:_delaun,boissonat86:_delaun_tree}. This is a
useful property, as we will now refine the initial partition by adding
higher-order periodic points to it.

\subsection{Refining the initial partition using orbits with a unique
    topological name}
\label{sec:partrefin1}

The three reference points and their associated symbols define an
initial partition.  However, this partition has a low precision and
cannot be reliably used except near one of the three reference points.
To refine it, we now have to extract information from the locations of
the higher-order periodic orbits.  To proceed as safely as possible,
we first consider the orbits which have a single topological name.

It should be noted that any cyclic permutation of the topological name
of an unambiguously identified orbit can in principle be used to label
its intersections with the section plane. Computing the Delaunay
triangulation of these periodic points, and determining the border as
explained above would yield a good partition, with different names
being given to different orbits.  Doing so, however, the border might
be so convoluted as to be useless because most points would be close
to the border. The description of such a partition would require an
enormous amount of information and the encoding of a chaotic
trajectory would be extremely sensitive to noise. It might also be
impossible to find a continuous encoding for the remaining orbits.

For each periodic orbit with a unique symbolic name, we thus have to
find the cyclic permutation of the symbolic name that keeps the
current partition as simple as possible. This can be achieved by
inserting periodic orbits in the partition in the following way.

Let us consider the next orbit beyond the period-$1$ and period-$2$
orbits, a period-$4$ orbit in our case (the attractor does not contain
period-$3$ orbits). This orbit is associated to two, possibly
different, symbolic names: (i) the topological name determined from
template analysis and (ii) the name that is obtained using the current
partition. Two situations may occur, depending on whether the latter
is a cyclic permutation of the former.

\begin{figure}[htbp]
  \begin{center}
    \leavevmode
    \includegraphics[width=12cm]{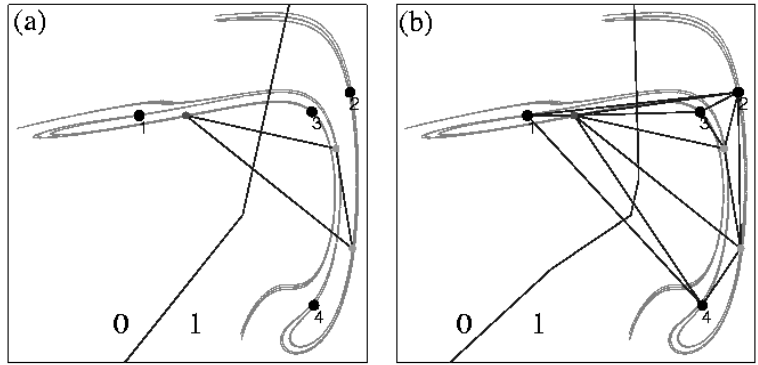}
    \caption{Inserting an orbit into the partition. (a) comparison of the
      topological name and of the name indicated by the current
      partition. In this case, the name obtained from the partition
      ($\pnameof{O_3}{\pa}=\perpoint{0111}$) matches the topological
      name. (b) the updated partition after the points of the
      period-$4$ orbit have been inserted into the list of reference
      points. Insertion of point \# 3 increases the precision of the
      partition. }
    \label{fig:addgood}
  \end{center}
\end{figure}

In the affirmative, the current partition correctly guesses the real
symbolic name of the orbit (Fig.~\ref{fig:addgood}a): we thus add its
points to the reference list, associated with the symbols indicated by
the current partition. If some of the new points are closer to the
border of the partition than the previous reference points, the
precision of the partition is increased (Fig.~\ref{fig:addgood}b).

If the topological and partition names of an orbit are not consistent,
we have to find the cyclic permutation of the topological name such
that the insertion in the triangulation of the corresponding pairs of
periodic points and symbols modifies the partition the least. To do
so, we determine for each permutation which of the periodic points
where the topological symbol differs from the one assigned by the
current partition is most distant from the partition border, and note
the corresponding distance. We then choose the cyclic permutation for
which this distance is the smallest, so that the border is displaced
by a small amount only.

\begin{figure}[htbp]
  \begin{center}
    \leavevmode
    \includegraphics[width=13cm]{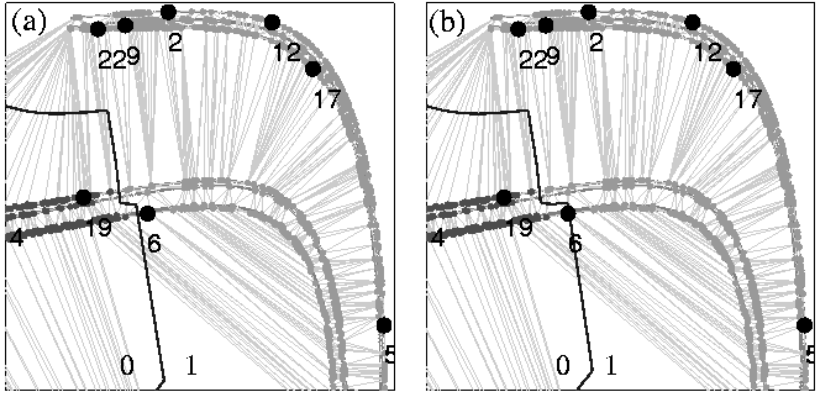}
    \caption{Comparison of the topological and partition names of the
      period-$23$ orbit $\orbit{(01)^2(011)^2(01011)^2(011)}$. (a) The
      symbolic name assigned by the current partition agrees with the
      topological name except at point \#6; (b) the updated partition
      after insertion of the orbit differs only slightly from the
      previous one, but point \#6 is now on the correct side of the
      border.}
    \label{fig:addbad}
  \end{center}
\end{figure}

A striking fact is that when carrying out the analysis of our sample
set of orbits, there was only one orbit, the period-$23$ orbit
$\orbit{(01)^2(011)^2(01011)^2(011)}$, for which the second rule had
to be used: the $249$ other orbits with a single topological name were
already correctly encoded by the partition under construction. This
orbit, and the partitions before after its insertion are shown in
Fig.~\ref{fig:addbad}.  It can be seen that the discrepancy is due to
a single point which is located very close to the border of the
current partition.

\begin{figure}[htbp]
  \begin{center}
    \leavevmode
    \includegraphics[angle=270,width=10cm]{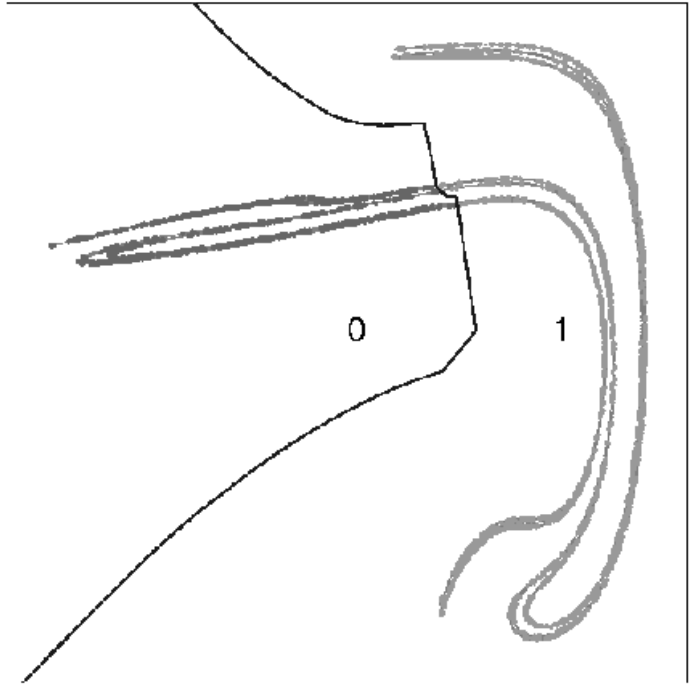}
    \caption{The partition as obtained from orbits with a
      single topological name. For clarity, the triangles are not
      shown. The large dots represent the reference points which
      parameterize the partition at this stage. It can be seen that
      the points for which the symbolic dynamical information can be
      unambiguously extracted cover well the attractor.  This provides
      a graphical illustration of the observation made about
      Table~\ref{tab:symnames}, but here with orbits of periods up to
      32. }
    \label{fig:partunamb}
  \end{center}
\end{figure}

After all orbits with a single topological name have been inserted, we
obtain a partition that: (i) assigns to each of these orbits its
topological name, (ii) has a simple structure, as can be seen in
Fig.~\ref{fig:partunamb}.

By using the fact that the border of this intermediate partition is
localized with a very good precision, we now proceed to the orbits for
which template analysis had selected several possible symbolic names,
and determine which of these names is the correct one. This will allow
us to further increase the resolution.

\subsection{Final stage of the construction}
\label{sec:partrefin2}

Periodic orbits with several topological names were not used in the
previous step, because we had then no reason of favoring one name over
the others. However, once an intermediate partition has been
determined from unambiguous orbits (Fig.~\ref{fig:partunamb}), it may
be used to determine the symbols of points that are far enough from
the border, if we assume that it will be only slightly modified by
further refinements.

\begin{figure}[htbp]
  \begin{center}
    \leavevmode
    \includegraphics[width=10cm]{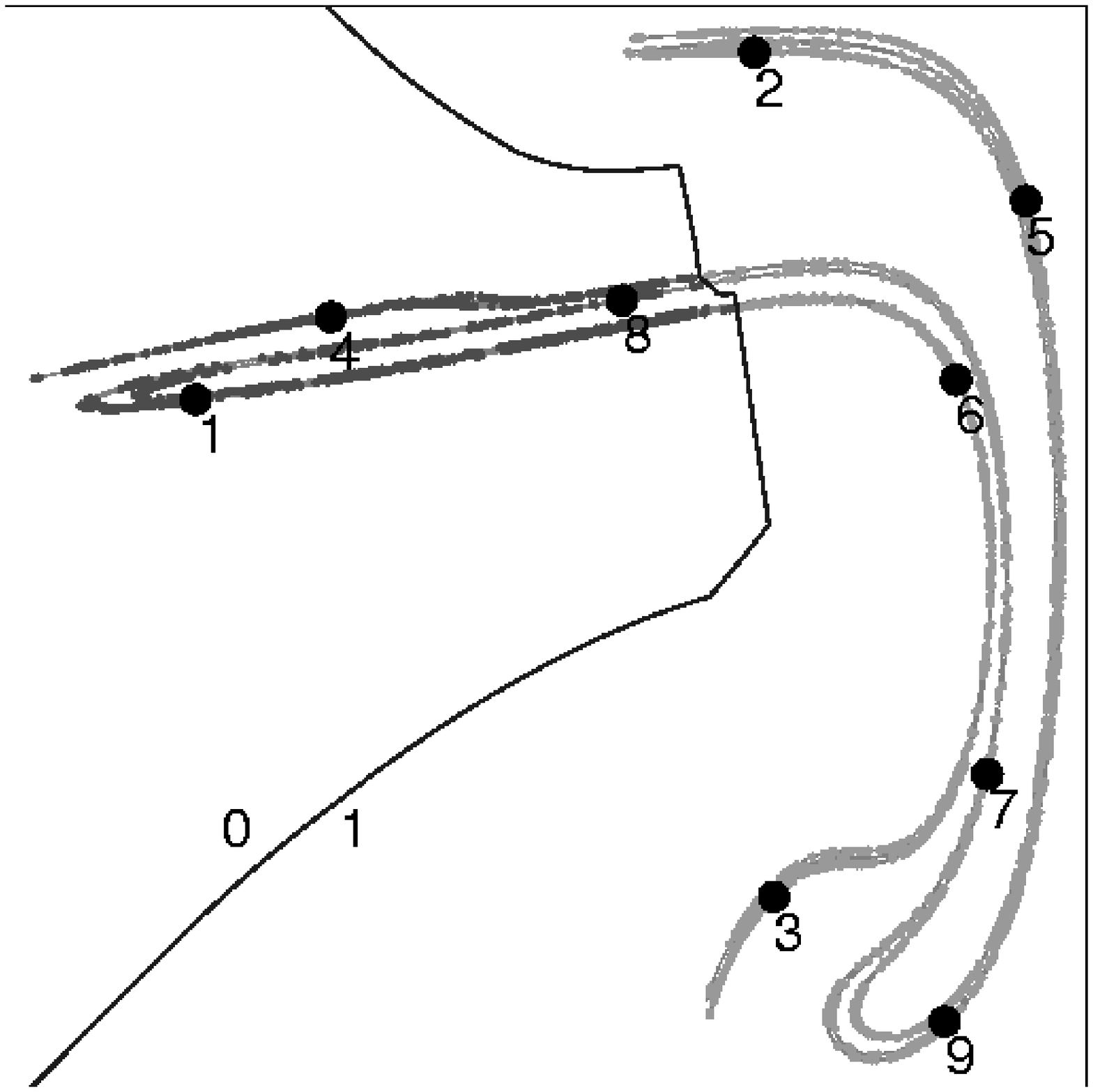}
    \caption{The periodic points of  orbit \#18 of Table
      \ref{tab:symnames}, which has two possible symbolic names, are
      represented with the partition obtained at the end of
      Sec~\ref{sec:partrefin1}. }
    \label{fig:choosing}
  \end{center}
\end{figure}

More precisely, consider the periodic orbit which is displayed in
Fig.~\ref{fig:choosing} (this is orbit \#18 of
Table~\ref{tab:symnames}). It has two possible names, namely
$\name_1=\orbit{010110111}=\orbit{0101^201^3}$ and
$\name_2=\orbit{010111011}=\orbit{0101^301^2}$. However, it can be
seen that its intersections with the section plane are far from the
partition border. Therefore, there is little doubt that the name
indicated by the current partition, which is
$\pname_{\pa}=\perpoint{01^301^201}$, is the correct one, as is
confirmed by the fact that it corresponds to a cyclic permutation of
$\name_2$. We can therefore assign this name to the orbit and insert
it in the partition. Then, by examining Table~\ref{tab:symnames}, one
immediately sees that since $\name_2$ has been assigned to orbit \#18,
it can no longer be a possible name for orbit \#19.  Therefore, the
only remaining possible name for the latter orbit is $\name_1$, which
indeed is also the one obtained from the current partition. We thus
see that a consistency check (different orbits should have different
names) allows us to identify the symbolic names of two orbits at once.

A more sophisticated consistency check that has to be carried after
the symbolic name of an orbit has been identified is whether all the
possible names of the not yet inserted orbits remain compatible with
the experimental table of topological invariants. Assume that, as in
the above example, the symbolic name of the $O_i$ orbit has just been
identified as being $\pname_i=\pnameof{O_i}{\pa}$. If a possible name
$\nameofi{O_j}{k}$ of another orbit $O_j$ is such that the linking
number $lk (\pname_i,\nameofi{O_j}{k})$ computed from the two names
does not match the measured value, $\nameofi{O_j}{k}$ can be discarded
without hesitation, as is illustrated in Table~\ref{tab:discard}. This
shows how enforcing simultaneously the requirements of smoothness and
of topological consistency allow one to solve the ambiguities
remaining after the template analysis step.

\begin{table}[htbp]
  \caption{Linking numbers of some Smale's horseshoe orbits.
    Assume that the names in  the first row (resp. first column) are
    the possible topological names of an experimental orbit $\alpha$
    (resp. $\beta$), and that the linking number of these two orbits
    is $lk(\alpha,\beta)=66$. If 
    the current partition can be used to show that the correct
    symbolic name of $\alpha$ can only be $\nameofi{\alpha}{2}$, then
    it follows immediately that the correct name for $\beta$ is
    $\nameofi{\beta}{2}$, since
    $lk(\nameofi{\alpha}{2},\nameofi{\beta}{1})$ does not match the
    experimentally measured invariant. }
  \label{tab:discard}
  \begin{tabular}[c]{|l|cc|}
    \hline\hline
    & $0101^201^3=\nameofi{\alpha}{1}$ &
    $0101^301^2=\nameofi{\alpha}{2}$ \\
    \hline
    $(01)^301^2(01^4)^2=\nameofi{\beta}{1}$ & 66 & 67 \\
    $(01)^3(01^4)^201^2=\nameofi{\beta}{2}$ & 67 & 66\\
    \hline\hline
  \end{tabular}
\end{table}

For some orbits, one or more periodic points are located in a close
neighborhood of the partition border. In this case, the name indicated
by the partition is uncertain: some symbols may not erroneous due to
the finite precision of the partition.  Yet, this provisional name can
be utilized to obtain the correct one, or at least to extract
additional information.  Indeed, if there are sequences of consecutive
periodic points $O_n^i, O_n^{i+1},\ldots,O_n^{i+k}$ whose symbols can
be determined unambiguously, this gives us substrings
$s_is_{i+1}\ldots s_{i+k}$ of the correct symbolic name of this orbit.
This information allows one to discard topological names that do not
contain this substring. If only one topological name remains, the
orbit can be inserted immediately in the partition. If there is still
an ambiguity, we delay the insertion of the orbit until further
information has been extracted from the other orbits.

The arguments presented above are very natural. Yet, to design a
precise algorithm, we must specify what ``far from the partition
border'' means. We thus need a precise rule to decide whether the
symbol assigned by the current partition to a given point $p$ in the
section plane can be trusted. We have found the following procedure to
be very reliable.

We first search for all the triangles of the current triangulation
whose circumcircle contains the point $p$, i.e. the triangles which
would be removed if $p$ was to be inserted in the
triangulation\footnote{Because a Delaunay triangulation has the
  property that the circumcircle of a triangle has no point in its
  interior, the insertion of a new point in a triangulation is
  performed by removing triangles whose circumcircle contains this
  point, and adding new ones so as to enforce the rule.}.  We then
examine the symbols associated with the vertices of these triangles.
If all these symbols are identical, we consider that the symbol
assigned to $p$ by the partition is certain.  If some symbols differ,
we conclude that the current partition is unreliable in the
neighborhood of $p$. The rationale of this rule is that insertion of a
point in the ``uncertain'' region defined in this way modifies the
border of the partition, because it modifies the triangles whose
circumcenters lie on the border.

There is however a small technical problem with this rule. Indeed it
is known that the outer edges of the triangulation of a set of points
compose the convex hull of this set. However, the support of the
strange attractor in the section plane is generally not convex because
of the folding process.  Consequently, there are triangles whose
vertices have different symbols merely because they are located on
opposite sides of the attractor (see, e.g., Fig.~\ref{fig:addbad}). A
direct use of the rule described above would then lead to conclude
that the symbol of points contained in the circumcircles of these
triangles cannot be reliably determined whereas the reference points
with different symbols are far away from each other on opposite sides
of the attractor.

\begin{figure}[htbp]
  \begin{center}
    \leavevmode
    \includegraphics[width=13cm]{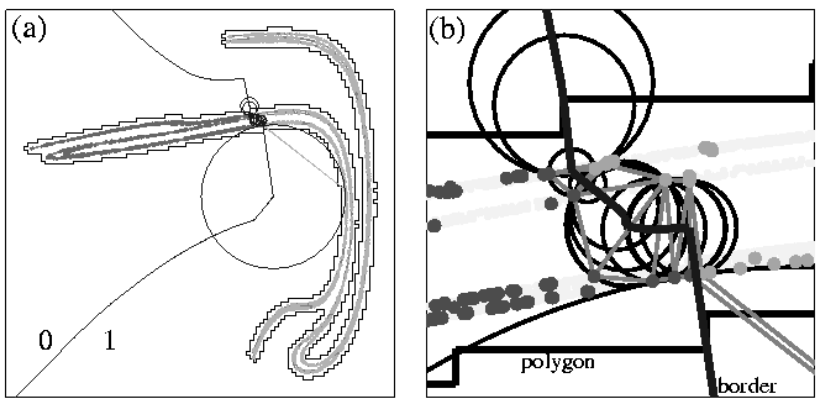}
    \caption{Regions of certain and uncertain coding. (a) a polygon
      providing a good approximation of the support of the strange
      attractor is determined. (b) Enlarged view of the border region.
      The uncertain region is defined to be located within the
      circumcircles of the triangles linking reference points with
      different symbols and whose mediators (which constitute the
      border line) lie within the shadow polygon.}
    \label{fig:uncertain}
  \end{center}
\end{figure}

To solve this difficulty by geometrical means, we compute a polygon
that tightly encloses the support of the attractor. Triangles with
different symbols are then classified according to whether the parts
of their mediators belonging to the partition border have a non-empty
intersection with the interior of the polygon, or not.  Only the first
class of triangles is used to assess the reliability of a symbol.
Thus, the modified rule states that the symbol of a point cannot be
reliably determined when the insertion of this point into the
triangulation would modify the partition border \emph{inside the
  support of the strange attractor}, which is illustrated in
Fig.~\ref{fig:uncertain}.

To summarize, the insertion of an orbit with several possible
topological names is carried out as follows. First, the symbolic
encoding of this orbit by the current partition is expressed by a
symbolic name $\pname_c$ with ``error bars''.  This symbolic name is
made of the symbols $\sym{0}$, $\sym{1}$, \ldots ,$\sym{n-1}$ (for
points that can be unambiguously coded) and $\sym{*}$ (for points
located in the ``uncertain'' region).  Then, we compare all cyclic
permutations of each topological name to this symbolic names, with
$\sym{*}$ matching any symbol. If two or more topological names are
compatible with $\pname_c$, we consider that we do not have enough
information at this point to insert the orbit, but nevertheless
discard the incompatible topological names.  On the contrary, if only
one topological name has a cyclic permutation that is compatible with
$\pname_{c}$, we consider that it is the correct symbolic name of the
orbit, and insert the orbit into the description of the partition.

\begin{figure}[htbp]
  \begin{center}
    \leavevmode
    \includegraphics[width=10cm]{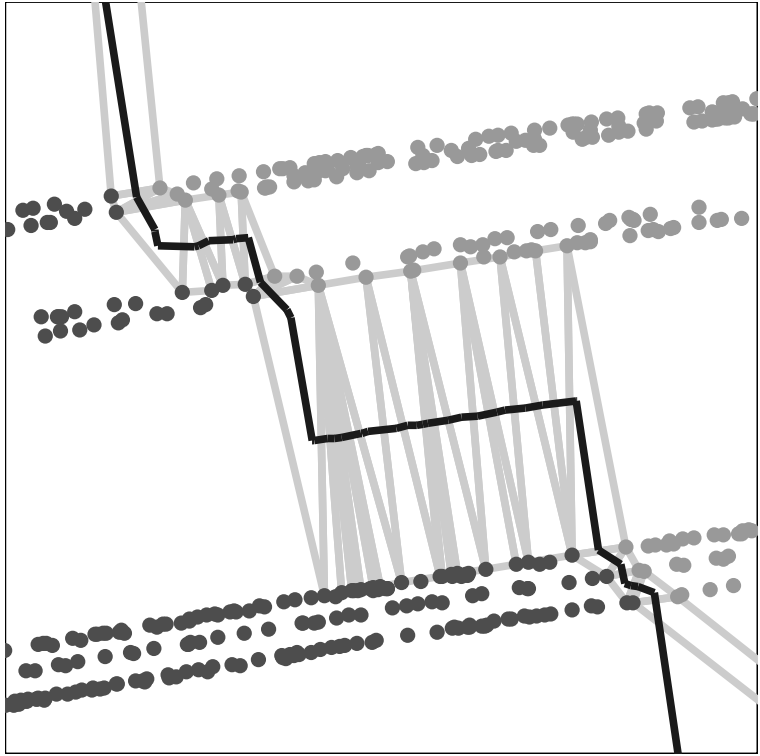}
    \caption{Enlarged view of the border of the final partition,
      obtained when all the periodic orbits have been inserted (the
      large-scale structure is virtually identical to that shown in
      Fig.~\ref{fig:partunamb}). The width of the represented box is
      $5\times 10^{-2}$ in units of the attractor width and the
      linewidth used to draw the border line and the reference points
      is $5\times 10^{-4}$.}
    \label{fig:finalpart}
  \end{center}
\end{figure}

Alternatively discarding names that are not compatible with the
current partition and names that are no longer compatible with
experimental topological invariants (as explained in
Table~\ref{tab:discard}) allows one to progressively insert all the
orbits, so that finally each orbit is associated with a single
symbolic name. The final partition, which is shown in
Fig.~\ref{fig:finalpart}, provides by construction a symbolic encoding
that is both consistent with the topological structure of the set of
periodic orbits and continuous (points that are close in the section
plane are associated to symbolic sequences that are close in the
symbol space).  Given the high number of periodic orbits in our
exemple, it is quite remarkable that the simple rules we have followed
naturally select a single name for each orbit: this supports the
existence of a well-defined symbolic encoding.

Note that the rule we have defined to assess the reliability of the
symbol associated to a point can be interpreted as an interpolation
problem. Indeed, Delaunay triangulations are routinely used to
interpolate the value of a function at an arbitrary point $p$ from
known values at the sites of the triangulation. There are essentially
two methods to do so. The first averages appropriately the values at
the three vertices of the triangle that contains $p$.  The second
utilizes the vertices of all the triangles whose circumcircle contains
$p$, and is called ``natural neighbor
interpolation''~\cite{watson85:_natur}.

Thus, it appears that our procedure, which we initially derived from
the heuristic argument presented above, is based on the latter.  When
the result of the interpolation is exactly one of the $n$ possible
symbols (because all neighboring vertices have the same symbol), the
result is considered as certain.  When the interpolation yields a
value that is intermediate between two symbols (some neighboring
vertices have different vertices), the encoding is considered as
uncertain. It is interesting to note that an earlier version of our
algorithm based on the first interpolation method did not converge in
some cases because some periodic points were incorrectly classified as
certain, leading to inconsistencies when inserting the remaining
orbits. This is because the reference point which is closest to $p$
need not be a vertex of the triangle containing $p$ whereas it is
known that it is a vertex of one of the triangles whose circumcircle
contains $p$.

In conclusion, it results that simple rules can be used to construct a
partition of the section plane using the information provided by (i)
the topological invariants of the unstable periodic orbits, and (ii)
their positions in the section plane. We have seen that this algorithm
yields partitions that have a very simple structure, and therefore
encode points with high reliability, except in a very small region
around the border of the partition.

\subsection{Increasing the resolution of the partition}
\label{sec:optim}

In describing our algorithm in the previous sections, our aim was to
show that no inconsistency was found even when shadowing all the
trajectories on the attractor with a high resolution, i.e., that it
was possible to assign unambiguously to every orbit a distinct
symbolic name compatible with its topological invariants. To this end,
we utilized a set of orbits that provided an uniform cover of the
attractor.  For practical applications, however, a high-resolution
cover is only needed in a small neighborhood of the border of the
partition.  To achieve a high precision at the lowest cost, we have
therefore modified our method as follows.

The procedure was split into two stages. First, an approximate
partition is determined using a set of orbits of limited period
providing a cover of moderate resolution. This allows one to bracket
the position of the border with reasonable accuracy. Using this
information, a second set of periodic orbits is selected so that it
provides a cover of the attractor with high resolution in the
neighborhood of the border, more precisely inside the circumcenters of
the triangles enclosing it, and with moderate resolution elsewhere.

We then apply to the latter set a slightly modified version of the
algorithm described in the previous sections. Indeed, we have
observed that obtaining a high-resolution cover in the critical region
requires using orbits of very high period, especially when there are
many forbidden sequences in the symbolic dynamics. This does not
induce additional difficulties in the first steps of the procedure
because (i) high-period orbits localized near the border of the
partition are marginally unstable, which makes their detection
relatively easy, (ii) topological invariants are expressed by integer
numbers and can therefore be reliably computed for orbits of very
large periods.

In fact, the limiting step is the search for possible symbolic names
using template analysis. Indeed, this search requires a considerable
amount of computing time because the number of symbolic names of
length $p$ increases exponentially with $p$, especially when the
symbolic dynamics is based on three or more symbols.  For a two-symbol
dynamics, the symbolic names of orbits with periods up to 32 can be
determined in a reasonable amount of computing time, while in the
three-symbol case, a direct search is practically limited to orbits of
period lower than 20.

We thus restrict this search to the orbits up to a certain period.  An
intermediate partition is built from these orbits, and is utilized to
list for the remaining higher-period orbits the symbolic names which
(i) are compatible with this partition, as explained in
Sec.~\ref{sec:partrefin2}, and (ii) correctly predict the topological
invariants of these higher-period orbits.

Once a list of possible names has been so obtained for each orbit in
the final set, the analysis proceeds as in Sec.~\ref{sec:partition}.
It should be stressed that this procedure is entirely equivalent to
the one described in Secs.~\ref{sec:partinit} to \ref{sec:partrefin2}
where all the topological names are determined before trying to build
the partition: we simply apply the selection criteria in a different
order.

\begin{figure}[htbp]
  \begin{center}
    \leavevmode
    \includegraphics[angle=270,width=13cm]{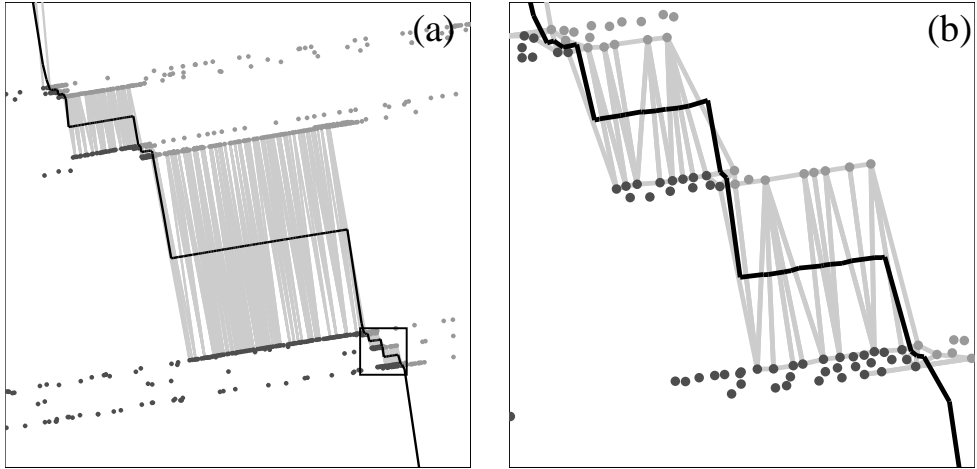}
    \caption{ (a) Enlarged view of the border of the partition
      obtained using a high-resolution cover of the border region by
      periodic points (the size of the represented box is the same as
      in Fig.~\ref{fig:finalpart}, i.e., $5\times 10^{-2}$ in units of
      the attractor width).  The triangles shown are those defining
      the partition border inside the shadow polygon.  Note the high
      density of periodic points associated with these triangles due
      to the design of our selection algorithm. (b) Enlarged view of
      the small square of size $5\times 10^{-3}$ displayed in the left
      picture. The linewidth used to draw the border line and the
      periodic points is $5\times 10^{-5}$. }
    \label{fig:highres_part}
  \end{center}
\end{figure}

With this modified algorithm, and a final set of 750 orbits of periods
up to 64, we have obtained for the chaotic attractor of
Fig.~\ref{fig:coding} a partition whose border is bracketed with a
resolution that is almost everywhere significantly below 0.01\% of the
attractor width (see Fig.~\ref{fig:highres_part}b). Note that the same
precision could be clearly be obtained at a lower cost by using a
significantly smaller number of periodic orbits. Indeed, it can be
seen in Fig.~\ref{fig:highres_part}a that many triangles connecting two
leaves of the attractor are in fact not essential for localizing
precisely the border, but were nevertheless considered by our
algorithm to belong to the border neighborhood. The many periodic
points associated with these triangles (the detection code is in
high-resolution mode in this region) could therefore be discarded from
the set of orbits without modifying the result.

It should be noted that such a precision is several orders of
magnitude higher than is needed for practical purposes. In fact, as we
will show in the second part of this work~\cite{PlumecoqP99b}, one has
to compare trajectories whose symbolic sequences have common
substrings of more than 60 symbols to observe an effect due to the
error in the location of the border.  Nevertheless, this example has
allowed us to verify the robustness of our algorithm down to very
small scales. Furthermore, it provides us with a test case which we
will use in the second part of this work to give evidence that our
approach is consistent with methods based on homoclinic tangencies.

\section{Conclusion and perspectives}
\label{sec:discussion_conclusion}

Our primary goal in showing that template analysis can be used to
obtain high-resolution partitions in numerical simulations was to give
strong evidence of the validity of the approach proposed
in~\cite{Lefranc94a}. A first result is the successful outcome of an
intensive check of the validity of template analysis: even when using
large sets of UPO of high periods, we could always find a global
projection on a simple branched manifold that preserves the
topological invariants.  Although the application of template theory
to real, nonhyperbolic, attractors is still lacking a rigorous
foundation, the present work gives further evidence that it accurately
describes the geometric structure of an attractor down to very small
scales. Although embedded in a nonhyperbolic attractor, the UPO appear
to be organized as in the hyperbolic limit, and constitute an
hyperbolic set which approximates well the strange attractor.

The foundation of this work is the fact that the knot invariants of an
unstable periodic orbit carry precise information about its symbolic
dynamics.  Building on this idea, we have described an algorithm to
construct generating partitions of a strange attractor. It combines
information (i) from the topological invariants of UPO embedded in it,
and (ii) from their location in a section plane, and is designed so as
to yield encodings that are continuous (sequences associated to
neighboring points should be close in symbol space). These basic
ingredients ensure that the resulting encodings are compatible with
those valid in the one-dimensional and hyperbolic limits, and that
they are dynamically relevant.

In our algorithm, a partition is described by a list of reference
points, whose associated symbols are given. To perform a symbolic
encoding, points in the section plane are associated with the symbol
carried by the closest reference point. This allows us to use simple
geometrical tools such as Delaunay triangulations.  Starting from an
initial configuration based on the lowest-period orbits, the accuracy
of the partition is progressively improved by adding periodic points
of increasing period to the list of reference points in a way that
preserves the simplicity of the partition and topological consistency.

Following this procedure, we have obtained partitions that have a
simple structure, yet reproduce the symbolic information extracted
from topological analysis: the unique symbolic name that is eventually
assigned to each periodic orbit is intimately related to its
topological structure, and hence to its
genealogy~\cite{Gilmore98a,Holmes88a,Mindlin93a,Hall93a}.  Had we
faced an inconsistency at some stage of the construction, we would
have been forced to conclude that there was a fundamental flaw in our
hypotheses. This was not the case neither in the example we considered
in this paper (we recall that it involved a set of 1594 periodic
orbits, whose topological information was contained in about
$4.4\times 10^6$ integer numbers), nor in others that we have
studied.

The present results call for further investigations in several
directions. First, we have to verify more extensively the relevance of
the obtained encodings, even if this should be guaranteed by the
consistency checks built into our algorithm. We do so in the second
part of this work~\cite{PlumecoqP99b}, where in particular we show
that the border of the high-resolution partition displayed in
Fig.~\ref{fig:highres_part} follows very accurately a line of
homoclinic tangencies, thus proving the equivalence of the two
approaches. We also give in~\cite{PlumecoqP99b} additional evidence of
the relevance of our algorithm by verifying (i) that encodings
obtained from different initial partitions are dynamically equivalent,
(ii) that accurate estimates of the metric entropy can be computed
from the probabilities of symbolic sequences, and (iii) that symbolic
sequences of increasing length select regions of decreasing diameter
in the section plane.

The robustness of our method with respect to noise should also be more
precisely studied. While template analysis behaves well in this
context, it would be desirable to quantify precisely the highest noise
level that is acceptable for extracting meaningful results. To achieve
this, a characterization of simulated time series contaminated by
various amounts of noise, and where UPO are detected from close
returns, is required.

Similarly, the independence of symbolic encodings with respect to
changes in parameter values should be carefully checked: in this work,
we have determined generating partitions only at a given set of
parameters. While it is obvious that the orbits with a unique
topological name will always be assigned the same name on their whole
domain of existence, we have to verify that this also holds for the
higher-period orbits whose identification is completed during the
construction of the partition.

This check will be absolutely required to be able to state with
reasonable confidence that the discontinuous changes in encodings
observed in orientation-reversing maps~\cite{Hansen92a,Giovannini92a}
cannot occur in orientation-preserving ones. The contradiction between
the phenomena reported in Refs~\cite{Hansen92a,Giovannini92a} and
constraints obeyed by orientation-preserving maps, which is unveiled
by simple topological arguments (see Sec.~\ref{sec:fund_hyp}),
certainly deserves further investigations in its own. In this
context, note that although knot invariants are defined for orbits of
a three-dimensional flow, the techniques described here can easily be
applied to invertible orientation-preserving two-dimensional maps,
such as the Ikeda map or the H\'enon map with positive Jacobian. One
can either construct a suspension of the map satisfying the uniqueness
theorem, or utilize the powerful techniques presented in
Refs.~\cite{Solari96b}, where it was shown that the braid
type of an orbit can be directly determined from its intersections
with a surface of section, up to a global torsion.

In the case of infinitely dissipative system, the relevant part of the
triangulation essentially consists of two periodic points located on
opposite sites of the border. In the mildly dissipative examples we
have considered in this work, the triangles enclosing the border
involve a significantly higher number of periodic points. It would be
highly interesting to determine whether these borderline periodic
points are directly related to the symbolic sequences defining the
``pruning front'' in symbol plane~\cite{Cvitanovic88a}. Another point
worth investigating is whether these orbits belong to a basis set (in
orbit forcing theory~\cite{Mindlin93a,Hall93a,Gilmore98a}, a basis set
is a small set of periodic orbits whose existence can be shown to
force the existence of all the other orbits embedded in the
attractor).

On the experimental side, we intend to apply very soon the algorithm
described here to a weakly dissipative experimental system, namely a
pump-modulated Nd:YAG laser. Indeed, the first experimental
topological encoding was obtained for a system which was relatively
dissipative (a CO$_2$ laser with modulated losses), even though its
return map could not be described by a well-defined one-dimensional
map on a wide range of parameters~\cite{Lefranc94a}.

To conclude, we would like to comment on the links that exist between
the topological approach we have discussed in this work and the
classical one based on homoclinic tangencies, although they seem to
have no common ground at first glance. As we will show
in~\cite{PlumecoqP99b}, these two methods yield results that are
equivalent, and thus must correspond to different views of a single
structure. In fact, both ultimately rely on the fact that a chaotic
invertible return map is a diffeomorphism organized by underlying
singularities.

To have these singularities appear undressed in the form of homoclinic
tangencies, it is necessary to iterate the return map an infinite
number of times. In doing so, however, one not only recovers the
organizing singularities, but also an infinite number of copies of
them. As a result, there is a fundamental ambiguity in the choice of
the homoclinic tangencies defining the partition. This can only be
solved by searching directly in the time-one map the singularities
that are hidden in it.

The Birman-Williams construction for hyperbolic systems provides a
deep answer to this problem. In the case of the horseshoe map, the
one-dimensional return map of the semi-flow defined on the branched
manifold features explicitly the fold singularity that is the
backbone of the two-dimensional horseshoe map, as well as any
three-dimensional suspension of it. This seems to indicate that the
structure of an invertible return map is inherited from a
lower-dimensional non-invertible map carrying the singularities that
organize the dynamics~\cite{Gilmore97b}.

Our approach does not try to construct directly such a singular map.
Rather, it focuses on the ``track'' of the one-dimensional
singularity, i.e., on points of the section plane that are projected
onto this singularity, if a ``projection'' similar to the
Birman-Williams one can be given a well-defined meaning.

The initial partition based on a few low-period orbits separates their
periodic point according to how they should located on an underlying
singular map, and in a sense provides a rough geometric modeling of
such a map. The hierarchical refinement of the partition carried out
by progressively inserting higher-order orbits can be viewed as a
means to ensure that ``iterates'' of this singular map converge to
iterates of the invertible return map. In this way,
the singularities which appear in the infinitely iterated return map
are eventually localized while keeping the geometrical description
close to that of the time-one map. This allows one to extract
information from the whole phase space without having to approach too
closely the noise-perturbed singularities.

This discussion makes it easier to understand why mathematical tools
that are intimately linked to the theory of one-dimensional symbolic
dynamics are so perfectly suited to the study of two-dimensional
dynamics. One key to this apparent paradox is that unstable periodic
orbits, among all other trajectories in the attractor, have a very
distinctive property: their forward symbolic sequence is uniquely
determined from their backward one, and vice versa. From a symbolic
dynamical point of view, periodic orbits are thus one-dimensional
objects intertwined with the fully two-dimensional chaotic
trajectories, which makes it possible to extend information extracted
from the former to the latter.

The bridge between the non-invertible and the invertible dynamics is
provided by the knot invariants of the UPO, which indicate how the
latter should be laid on the domain of the underlying one-dimensional
map without having to construct it explicitly.  These invariants thus
play a role that is not unlike the conserved quantities or symmetries
that have proved so immensely useful in many fields of physics. To
extend topological coding to higher-dimensional chaos, and understand
its singularity structure, we thus now have to find what are in this
case the appropriate invariant quantities.

\begin{ack}
  It is a pleasure to thank our colleagues Guillaume Boulant, Serge
  Bielawski, Dominique Derozier, and Robert Gilmore for stimulating
  discussions.  The Laboratoire de Physique des Lasers, Atomes,
  Mol\'ecules is Unit\'e de Recherche Mixte du CNRS. The Centre
  d'\'Etudes et Recherches Lasers et Applications is supported by the
  Minist\`ere charg\'e de la Recherche, the R\'egion Nord-Pas de
  Calais and the Fonds Europ\'een de D\'eveloppement \'Economique des
  R\'egions.
\end{ack}

\end{document}